\documentclass[article]{jss}

\usepackage{orcidlink,thumbpdf,lmodern}
\usepackage{amsmath, amssymb, booktabs}

\newcommand{\fct}[1]{\code{#1()}}

\DeclareMathOperator\avar{avar}

\author{Yuhao Deng~\orcidlink{0000-0003-0331-6070}\\ Fred Huchinson Cancer Center
   \And Yi Zhou~\orcidlink{0000-0001-9254-3245}\\ Kobe University
   \AND Yi Wang~\orcidlink{0000-0003-0253-1334} \\ Shanghai University of \\ International Business \\ and Economics
   \And Shasha Han~\orcidlink{0000-0001-7388-8125}\\ Chinese Academy of \\ Medical Sciences \& Peking \\ Union Medical College
   }
\Plainauthor{Yuhao Deng, Yi Zhou, Yi Wang, Shasha Han}

\title{\pkg{tteICE}: An \proglang{R} Package for Estimating Treatment Effects on Time-to-Event Data with Intercurrent Events in Two-Arm Trials}
\Plaintitle{tteICE: An R Package for Estimating Treatment Effects on Time-to-Event Outcomes with Intercurrent Events in Two-Arm Trials}
\Shorttitle{tteICE in \proglang{R}}

\Abstract{
Evaluating treatment effects in clinical trials with time-to-event outcomes is complicated by intercurrent events (ICEs), e.g., treatment discontinuation or competing events. While the ICH E9 (R1) addendum outlines five strategies to address these challenges, accessible software implementing these methods is limited. This article introduces the \pkg{tteICE} package for \proglang{R}, which implements five strategies to facilitate valid causal inference in the presence of ICEs. \pkg{tteICE} enables researchers to either nonparametrically or semiparametrically efficiently estimate potential cumulative incidence functions under both treatment and control conditions. The treatment effect is assessed by contrasting these functions, with statistical uncertainty quantified using pointwise confidence intervals and $p$-values. The package supports data from randomized controlled trials and observational studies, including settings with competing risks and semi-competing risks. To reduce barriers for applied researchers, \pkg{tteICE} features integrated plotting functions and an accompanying interactive Shiny application that provides a user-friendly graphical interface with step-by-step instructions. 
}

\Plainkeywords{Time-to-event outcome, intercurrent event, survival analysis,   cumulative incidence function,  R}

\Address{
  Yi Zhou\\
  Division of Mathematics and Informatics\\ 
  Graduate School of Human Development and Environment \\
  Kobe University\\
  3-11 Tsurukabuto, Nada-ku \\
  Kobe 657-8501, Japan\\
  E-mail: \email{zhouy@people.kobe-u.ac.jp}
}

\begin{document}

\section[Introduction]{Introduction} \label{sec:intro}

The analysis of time-to-event (or survival) data is a cornerstone of medicine and health research. The statistical community has developed a rich suite of software tools for this purpose. 
In \proglang{R} \citep{R}, packages like \pkg{survival} \citep{survival-package} provide core functionality for nonparametric estimation (e.g., \fct{survfit} for nonparametric Kaplan--Meier curves, \fct{survdiff} for testing the difference in survival functions) and semiparametric estimation (e.g., \fct{coxph} for the Cox proportional hazards model). While the proportional hazards model is the most widely used semiparametric model due to its interpretability and estimation efficiency, alternative semiparametric models are also implemented. For example, packages like \pkg{timereg} \citep{scheike2011analyzing} (in which \fct{aalen} and \fct{additive.regression} for additive hazards models \citep{martinussen2006dynamic}) and \pkg{flexsurv} \citep{jackson2016flexsurv} (in which \fct{flexsurvreg} for parametric regression and \fct{flexsurvspline} for B-spline models) offer more advanced methods.

A significant complication in practical trials is the occurrence of intercurrent events (ICEs), which are post-randomization events such as treatment discontinuation or rescue medication that affect the interpretation or measurement of the primary outcome. For example, to study a drug's effect on time to heart failure, any microvascular event is an ICE that renders the primary outcome uninterpretable, and non-cardiovascular death is an ICE that renders the primary outcome unmeasurable. Traditional survival analysis often treats ICEs as censoring events or incorporates them into competing risks, semi-competing risks, or multi-state models \citep{gray1988class, fine1999proportional, fine2001semi, lee2017accelerated}. The estimand is usually defined on the scale of the cumulative incidence function (CIF) of the primary outcome rather than the survival function in the presence of ICEs, because the CIF is always well-defined, regardless of whether the primary outcome can be interpreted or measured. 

In \proglang{R}, these models are typically implemented in packages such as \pkg{cmprsk} \citep{gray2010cmprsk}, \pkg{fastcmprsk} \citep{kawaguchi2020fast}, \pkg{tidycmprsk} \citep{tidycmprsk}, \pkg{mets} \citep{holst2016liability} for competing risks, \pkg{SemiCompRisks} \citep{alvares2019semicomprisks} for semi-competing risks, \pkg{survival} \citep{survival-package}, \pkg{flexsurv} \citep{jackson2016flexsurv} and \pkg{mstate} \citep{de2010mstate, de2011mstate}, \pkg{SemiMarkov} \citep{krol2015semimarkov} for multi-state models.  
In \proglang{SAS}, the estimation of the CIF is primarily handled by two procedures: \code{PROC LIFETEST} and \code{PROC PHREG} \citep{lin2012analyzing, guo2018cause}. In \proglang{STATA}, commands \code{stcompet} \citep{coviello2004cumulative}, \code{stcrreg} \citep{gutierrez2010competing}, \code{stpm2cif} \citep{hinchliffe2013extending}, and \code{stpm2cr} \citep{mozumder2017flexible} estimate the CIF for competing risks, \code{stpm2illd} \citep{hinchliffe2013flexible} for semi-competing risks, and package \pkg{multistate} \citep{crowther2023multistate} for multi-state models.
Although these packages facilitate the implementation of various statistical methods for survival analysis, they do not provide formal guidelines for users to conduct causal analyses to estimate treatment effects in the presence of ICEs. 

The ICH E9 (R1) addendum \citep{ICH19} provides a crucial framework for addressing this challenge by outlining five strategic principles (treatment policy, composite, while on treatment, hypothetical, and principal stratum) to define causal estimands that incorporate ICEs. ICH E9 (R1) was initially proposed in the context of randomized controlled trials (RCTs); however, these five strategies can be universally generalized to observational studies by appropriately adjusting for confounding. 
Among the strategies, some can be awkwardly implemented using existing software, requiring substantial data preprocessing (e.g., the use of competing risks models for the while on treatment strategy and \textit{ad hoc} data handling for the composite strategy); thus, a significant software gap exists for non-specialist users. 
Even within the widely used \proglang{R} ecosystem, no package is designed to address ICEs from a general causal perspective, and existing packages address only specific questions. For example, the \proglang{R} \pkg{speff2trial} package \citep{juraska2022package} implements semiparametrically efficient estimation methods in two-arm trials without addressing intercurrent events, the \pkg{CausalCmprsk} package \citep{vakulenko2023causalcmprsk} implements propensity score weighting methods for competing risks data, the \pkg{mediation} package \citep{tingley2014mediation} incorporates time-to-event data into a wide range of mediation analysis settings, and the \pkg{PStrate} package \citep{liu2024principal} targets a principal stratum treatment effect.
To date, no unified tool directly implements these strategies within a formal causal inference framework, especially the computationally demanding hypothetical and principal stratum approaches.

This paper introduces the \pkg{tteICE} package for \proglang{R}, designed to fill this critical gap in the software ecosystem. Although most strategies in ICH E9 (R1) are formalized in randomized clinical trials \citep{deng2025inference}, our package operationalizes and extends these strategies to observational studies to estimate causal treatment effects. In such settings, the assumption of unconfoundedness (or exchangeability) is critical and typically relies on the inclusion of a comprehensive set of pre-treatment covariates. Our package provides a systematic implementation of the ICH E9 (R1) framework, enabling researchers to estimate potential cumulative incidence functions and treatment effects for time-to-event outcomes in the presence of ICEs. It supports both competing and semi-competing risks data structures. The package offers nonparametric, weighting-based, and semiparametrically efficient estimators, together with statistical inference (pointwise confidence intervals and $p$-values) and integrated visualization tools. To maximize accessibility and usability, particularly for clinical practitioners, we complement the core \proglang{R} package with an interactive Shiny application that provides a graphical interface and step-by-step guidance for the entire analysis workflow.

The remainder of this paper is organized as follows. Section \ref{sec:methods} provides a background of statistical methods for addressing ICEs. Section \ref{sec:implement} introduces the software architecture and the detailed implementations of its functions. Section \ref{sec:use} introduces high-level functions that aggregate the methods, which users can simply call in practice. Section \ref{sec:illust} demonstrates the functionalities of this package using a practical example. Finally, we conclude with a discussion in Section \ref{sec:summary}. Technique details are postponed to Appendix \ref{sec:technical}, and simulation studies are given in Appendix \ref{sec:simulation}.

\section{Five strategies to address intercurrent events} \label{sec:methods}

According to the ICH E9 (R1) addendum, estimand strategies should be selected based on the clinical question of interest prior to data analysis. 
An overview of the five strategies and the corresponding questions to answer is summarized in Table \ref{tab:ICH}, contextualized within common clinical research scenarios. The methodological descriptions and analyses presented in the subsequent sections are primarily framed within the context of RCTs to establish a clear causal foundation for the estimands. In practice, the analytical framework is also applicable to observational data under specific conditions in which a target trial can be explicitly emulated (see Section \ref{sec:implement} for details).

\begin{table}[tbh]
\centering
\caption{A practical guide to ICH E9 (R1) estimand strategies for time-to-event outcomes}\label{tab:ICH}
\begin{tabular}{p{2.1cm} p{6cm} p{5.7cm}}
\toprule
\textbf{Strategy} & \textbf{The scientific question to answer} & \textbf{Key considerations or assumptions} \\
\midrule
Treatment policy (tp) &
What is the effect of \textit{assigning} the treatment, allowing ICEs to occur naturally? &
ICEs do not prevent the observation of the primary outcome event. It reflects the real-world effectiveness of the initial assignment. \\
Composite variable (cv) &
What is the effect on the time to the \textit{first clinically meaningful event}, either the primary outcome event or ICEs? &
The ICE is considered a clinically meaningful outcome. The composite event must be interpretable. \\
While on treatment (wo) &
What is the effect on the primary outcome event while remaining \textit{free from} ICEs? &
It evaluates the cause-specific risk. Interpretability can be compromised if treatment arms have very different risks of ICEs. \\
Hypothetical (hp) &
What \textit{would} the effect on the primary outcome event be \textit{if} the hazard of ICEs \textit{were} controlled (e.g., set to its level in the control group, or completely removed)? &
A hypothetical scenario is envisioned. Causal assumptions and statistical modeling are required for interpretation (e.g., sequential ignorability or Markovianity). \\
Principal stratum (ps) &
What is the effect specifically in the \textit{subgroup} of patients (e.g., who would \textit{never} experience ICES), regardless of their treatment assignment status? &
The target population is unobservable and must be inferred. The principal ignorability assumption is required. \\
\bottomrule
\end{tabular}
\end{table}

\subsection{Settings}

Consider a randomized controlled trial with $n$ individuals randomly assigned to one of two treatment conditions, denoted by $w$, where $w = 1$ represents the active treatment (a test drug) and $w = 0$ represents the control (placebo). 
An upper limit on the study duration, starting from treatment initiation, is set to $t^*$. In practice, $t^*$ is typically set to the maximum follow-up time.
Assume that all patients adhere to their treatment assignments and do not discontinue treatment before the end of the study. For illustration, we assume there is at most one intercurrent event. Otherwise, we can focus on the first intercurrent event. 
As such, for each individual, there are two potential outcomes, $T(1)$ and $T(0)$, representing the durations from treatment initiation to the primary outcome event under the two treatment assignments, respectively, whenever these durations are well-defined. 
Let $R(1)$ and $R(0)$ denote the occurrence time of potential intercurrent events, if any, under the two treatment assignments, respectively. If primary outcome events are prevented by intercurrent events under treatment condition $w$, we denote $T(w) = \infty$. 
Intercurrent events are considered as absent if no post-treatment intercurrent events occur until $t^*$, in which case we denote $R(w) = \infty$ or an arbitrary number larger than $t^*$, since events after $t^*$ will not be counted in the analysis.

In the following, we use ${P}(\cdot)$ to denote the probability taken with randomized sampling and treatment assignment, and $I(\cdot)$ to denote the indicator function. 
Each individual is associated with two potential right-censoring times, $C(1)$ and $C(0)$. Because of censoring, we are only able to partially observe potential primary outcome events through $\Delta^T(w) = I\{T(w) \leq C(w)\}$ and $\tilde{T}(w) = T(w) \wedge C(w)$, and potential intercurrent events through $\Delta^R(w) = I\{R(w) \leq T(w) \wedge C(w)\}$ and $\tilde{R}(w) = R(w) \wedge T(w) \wedge C(w)$. 
Let $\tilde{T}$ be the observed time to the primary outcome event (or censoring) with event indicator $\Delta^T$, and $\tilde{R}$ be the observed time to the intercurrent event (or censoring) with event indicator $\Delta^R$. 
We make the following assumptions, which are standard in causal inference and survival analysis:
\begin{itemize}
\item Stable unit treatment value assumption (SUTVA): There is only one version of treatment and one version of control; there is no interference between units.
\item Randomization: The treatment assignment $W$ is independent of all potential outcomes.
\item Random censoring: The potential censoring time $C(w)$ is independent of $(T(w),R(w))$.
\item Causal consistency: $\tilde{T}=\tilde{T}(W)$, $\tilde{R}=\tilde{R}(W)$, $\Delta^T=\Delta^T(W)$, and $\Delta^R=\Delta^R(W)$, so that potential outcomes associated with the received treatment condition are observable.
\item Positivity: There is a positive probability of still being at risk at the end of the study in each group.
\end{itemize}

The observed data structure can be simplified according to how intercurrent events compete with primary outcome events. If intercurrent events do not prevent the occurrence of the primary outcome events (semi-competing risks data), our observed data structure takes the form $(W, \tilde{T}, \tilde{R}, \Delta^T, \Delta^R)$. On the other hand, if intercurrent events can prevent the occurrence of primary outcome events (competing risks data), our observed data structure takes the form $(W, \tilde{T} \wedge \tilde{R}, \Delta^T, \Delta^R)$, since the individual will only be followed until the first occurrence of any event. For competing risks data, the event indicators $\Delta^T$ and $\Delta^R$ are usually recoded as a single value $J$, where $J = 1$ if the primary outcome event is observed, $J = 2$ if the intercurrent event is observed, and $J = 0$ if the censoring event is observed. Semi-competing risks data can be transformed into competing risks data by keeping the earlier event with its corresponding indicator.
We assume there are no ties (i.e., the intercurrent and primary outcome events cannot occur simultaneously); if we observe the same time of occurrence for the intercurrent and primary outcome events, we assume the intercurrent event occurs slightly before the primary outcome event.

\subsection{Estimands}

Due to intercurrent events, the times to potential primary outcome events, $T(1)$ and $T(0)$, may be ill-defined or not clinically meaningful, as they reflect a mixed effect of treatment assignment and intercurrent events. Therefore, we must carefully choose the causal estimand. 
Let $\mu_1^{k}(t)$ and $\mu_0^{k}(t)$ ($k \in \{\text{tp, cv, hp, wo, ps}\}$, $0 \leq t \leq t^*$) denote the two cumulative incidence functions of potential outcome events under the treatment policy strategy ($k=\text{tp}$), composite variable strategy ($k=\text{cv}$), while on treatment strategy ($k=\text{wo}$), hypothetical strategy ($k=\text{hp}$) and principal stratum strategy ($k=\text{ps}$), respectively. The corresponding causal estimands are defined as $\tau^k(t)=\mu_1^{k}(t)-\mu_0^{k}(t)$ with further details introduced below.

\paragraph{Treatment policy strategy.}

The treatment policy strategy addresses the problem of ICEs by expanding the initial treatment conditions to a treatment policy $(w,R(w))$, which includes the initial treatment condition and ICEs that develop naturally. Therefore, the potential time to the primary outcome event under the treatment policy $(w,R(w))$ is reexpressed as $T(w,R(w))$. This strategy is applicable only if ICEs do not prevent primary outcome events. It is confined to the semi-competing risks data structure. 
Rather than comparing the test drug and placebo directly, the contrast of interest is between the two treatment policies. The difference in cumulative incidence functions under the two treatment policies is then
\begin{equation}\label{esmd:diff:tp}
\begin{aligned}
\tau^{\text{tp}}(t) &:= \mu_1^{\text{tp}}(t) - \mu_0^{\text{tp}}(t) \\
&:= {P}(T(1, R(1)) < t) - {P}(T(0, R(0)) < t),
\end{aligned}
\end{equation}
representing the difference in probabilities of experiencing primary outcome events during $[0, t)$ under active treatment and placebo.
The average treatment effect $\tau^{\text{tp}}(t)$ has a meaningful causal interpretation only when $T(1, R(1))$ and $T(0, R(0))$ are well-defined. Because the treatment policy treats the occurrence of the intercurrent event as natural, the entire policy is determined solely by the initial treatment condition $w$. Therefore, we can simplify the notations $T(w, R(w)) = T(w)$ in defining estimands with $w \in \{1, 0\}$. Accordingly, $\tau^{\text{tp}}(t) = {P}(T(1) < t) - {P}(T(0) < t)$ corresponds to the intention-to-treat estimand. Only the time to the primary outcome event and the event indicator are needed to implement this strategy.

\paragraph{Composite variable strategy.}

The composite variable strategy addresses the problem of ICEs by expanding the outcome variables. It aggregates the ICEs and the primary outcome event into a single composite outcome variable. 
One widely used composite outcome variable has the form $R(w) \wedge T(w) = \min\{T(w), R(w)\}$ for $w \in \{1,0\}$. The difference in counterfactual cumulative incidence functions is
\begin{equation}\label{esmd:diff:cv}
\begin{aligned}
\tau^{\text{cv}}(t) &:= \mu_1^{\text{cv}}(t) - \mu_0^{\text{cv}}(t) \\
&:= {P}( R(1) \wedge T(1) < t ) - {P}( R(0) \wedge T(0) < t ),
\end{aligned}
\end{equation}
representing the difference in probabilities of experiencing either ICEs or primary outcome events during $[0, t)$ under active treatment and placebo. Since only the minimum of $T(w)$ and $R(w)$ is used, the composite variable strategy can be applied to both competing risks and semi-competing risks data.

\paragraph{While on treatment strategy.}

The while on treatment strategy considers the measurement of outcome variables only up to the occurrence of ICEs. Primary outcome events should not be counted in the cumulative incidence functions if ICEs have occurred. The difference in cumulative incidence functions under this strategy is
\begin{equation}\label{esmd:diff:wo2}
\begin{aligned}
\tau^{\text{wo}}(t) &:= \mu_1^{\text{wo}}(t) - \mu_0^{\text{wo}}(t) \\
&:= {P}(T(1) < t, R(1) \geq t) - {P}(T(0) < t, R(0) \geq t),
\end{aligned}
\end{equation}
representing the difference in probabilities of experiencing primary outcome events without intercurrent events during $[0, t)$ under active treatment and placebo. The $\mu_w^{\text{wo}}(t)$ is also known as the cause-specific cumulative incidence or subdistribution function \citep{fine1999proportional, lau2009competing}.
The while on treatment strategy is closely related to the competing risks model. We can rewrite ${P}(T(w)<t, R(w)\geq t) = {P}(T(w) \wedge R(w)<t, T(w)<R(w))$; that is, the first event occurs before $t$, and this event is the primary outcome event. Since only the minimum of $T(w)$ and $R(w)$ is used, the while on treatment strategy can be applied to both competing risks and semi-competing risks data.
However, for causal interpretations, it is worth emphasizing that the hazard of $R(1)$ may differ from that of $R(0)$, leading to a vast difference in the underlying features of individuals who have not experienced the primary outcome event between treatment conditions at any time $t \in [0,t^*)$. When the scientific question of interest is the impact of treatment on the primary outcome event, the estimand $\tau^{\text{wo}}(t)$ is hard to interpret if a systematic difference in the risks of intercurrent events between two treatment conditions under comparison is anticipated.

\paragraph{Hypothetical strategy.}

The hypothetical strategy envisions a hypothetical clinical trial condition in which the occurrence of ICEs is restricted in certain ways. By doing so, the distribution of potential outcomes in the hypothetical scenario can explicitly capture the impact of intercurrent events through a pre-specified criterion. 
We use $T'(w)$ to denote the time to the primary outcome event in the hypothetical scenario. The time-dependent treatment effect specific to this hypothetical scenario is written as
\begin{equation}\label{esmd:diff:hp}
\begin{aligned}
\tau^{\text{hp}}(t) &:= \mu_1^{\text{hp}}(t) - \mu_0^{\text{hp}}(t) \\
&:= {P}(T'(1) < t) - {P}(T'(0) < t),
\end{aligned}
\end{equation}
representing the difference in probabilities of experiencing primary outcome events during $[0, t)$ in the pre-specified hypothetical scenario under active treatment and placebo.
The key question is how to envision $T'(1)$ and $T'(0)$. For the competing risks data structure, we may manipulate the hazard specific to ICEs $d\Lambda_2(t; w)$ while assuming the hazard specific to the primary outcome event $d\Lambda_1(t; w) = {P}(t \leq T(w) < t+dt \mid T(w) \geq t, R(w) \geq t)$ remains unchanged, where $w \in \{1, 0\}$ and $0 \leq t \leq t^*$. 
Let $d\Lambda_2'(t;w)$ and $d\Lambda_1'(t;w)$ be the hazards specific to ICEs and the primary outcome event in the hypothetical scenario, respectively, with $d\Lambda_1'(t;w) = d\Lambda_1(t;w)$. 

Specifically, we give two estimands under the hypothetical strategy. The first corresponds to a scenario in which ICEs that occur under assignment to the test drug are allowed only if the same ICEs would also have occurred had the individuals been assigned to a placebo. In this hypothetical scenario, when assigned to a placebo, individuals would be equally likely to experience ICEs as they are assigned to a placebo in the real-world trial in terms of the hazards; when assigned to the test drug, the hazard of ICEs would be identical to that if assigned to a placebo in the real-world trial. That is, $d\Lambda_2'(t;0) = d\Lambda_2'(t;1) = d\Lambda_2(t;0)$. We refer to this as hypothetical scenario I (interpreted as the natural effect in mediation analysis) and denote the corresponding estimand by $\tau^{\text{hp,I}}(t) = \mu_1^{\text{hp,I}}(t)-\mu_0^{\text{hp,I}}(t)$. This effect can also be interpreted as separable effects under some causal assumptions \citep{martinussen2023estimation}.
Alternatively, we can envision another hypothetical scenario where ICEs are absent in the hypothetical scenario for all individuals, so $d\Lambda_2'(t;0) = d\Lambda_2'(t;1) = 0$. We refer to this as hypothetical scenario II (interpreted as the controlled effect with ICEs removed in mediation analysis) and denote the corresponding estimand by $\tau^{\text{hp,II}}(t) = \mu_1^{\text{hp,II}}(t)-\mu_0^{\text{hp,II}}(t)$. This hypothetical scenario II leads to an estimand called the marginal cumulative incidence function.

The hypothetical strategy can also be applied to semi-competing risks data by modifying how the hypothetical scenarios are envisioned. We still consider the two hypothetical scenarios mentioned above. In scenario I, we assume the hazard of ICEs is identical under the test drug and a placebo. Primary outcome events after ICEs are allowed. Note that there are two cause-specific hazards for the primary outcome event: one without ICEs and the other with ICEs. The cause-specific hazards of the primary outcome event are allowed to vary with treatment conditions. In scenario II, we assume the hazard of ICEs is zero. The cause-specific hazard of the primary outcome event associated with ICEs would no longer matter, as ICEs are removed. The estimand in this scenario will be identical to that using competing risks data. However, it is worth noting that while envisioning hypothetical scenarios does not require causal assumptions, interpreting the estimands as natural or controlled effects in mediation analysis requires sequential ignorability, i.e., that there is no unmeasured confounding between ICEs and the primary outcome event \citep{deng2024direct}.

\paragraph{Principal stratum strategy.}

The principal stratum strategy aims to stratify the population into subpopulations based on the joint potential occurrences of ICEs $(R(1), R(0))$ under the two treatment assignments. Suppose we are interested in a principal stratum comprising individuals who would never experience ICEs, regardless of the treatment they receive. This principal stratum can be indicated by $\{R(1)=R(0)=\infty\}$. The treatment effect is now defined within this subpopulation,
\begin{equation}\label{esmd:diff:ps}
\begin{aligned}
\tau^{\text{ps}}(t) &:= \mu_1^{\text{ps}}(t) - \mu_0^{\text{ps}}(t) \\
&:= {P}(T(1) < t \mid R(1)=R(0)=\infty) 
- {P}(T(0) < t \mid R(1)=R(0)=\infty),
\end{aligned}
\end{equation}
representing the difference in probabilities of experiencing primary outcome events during $[0, t)$ under active treatment and placebo in the subpopulation that will not experience ICEs regardless of treatment during $[0, t)$.
Identification of this estimand relies on untestable assumptions, typically principal ignorability, saying that $T(w) \perp R(1-w) \mid R(w)$, for $w = 1, 0$. Since only the minimum of $T(w)$ and $R(w)$ is used, the composite variable strategy can be applied to both competing risks and semi-competing risks data.

\section{Implementation of methods} \label{sec:implement}

\begin{figure}[tbh]
\centering
\includegraphics[width=0.9\textwidth]{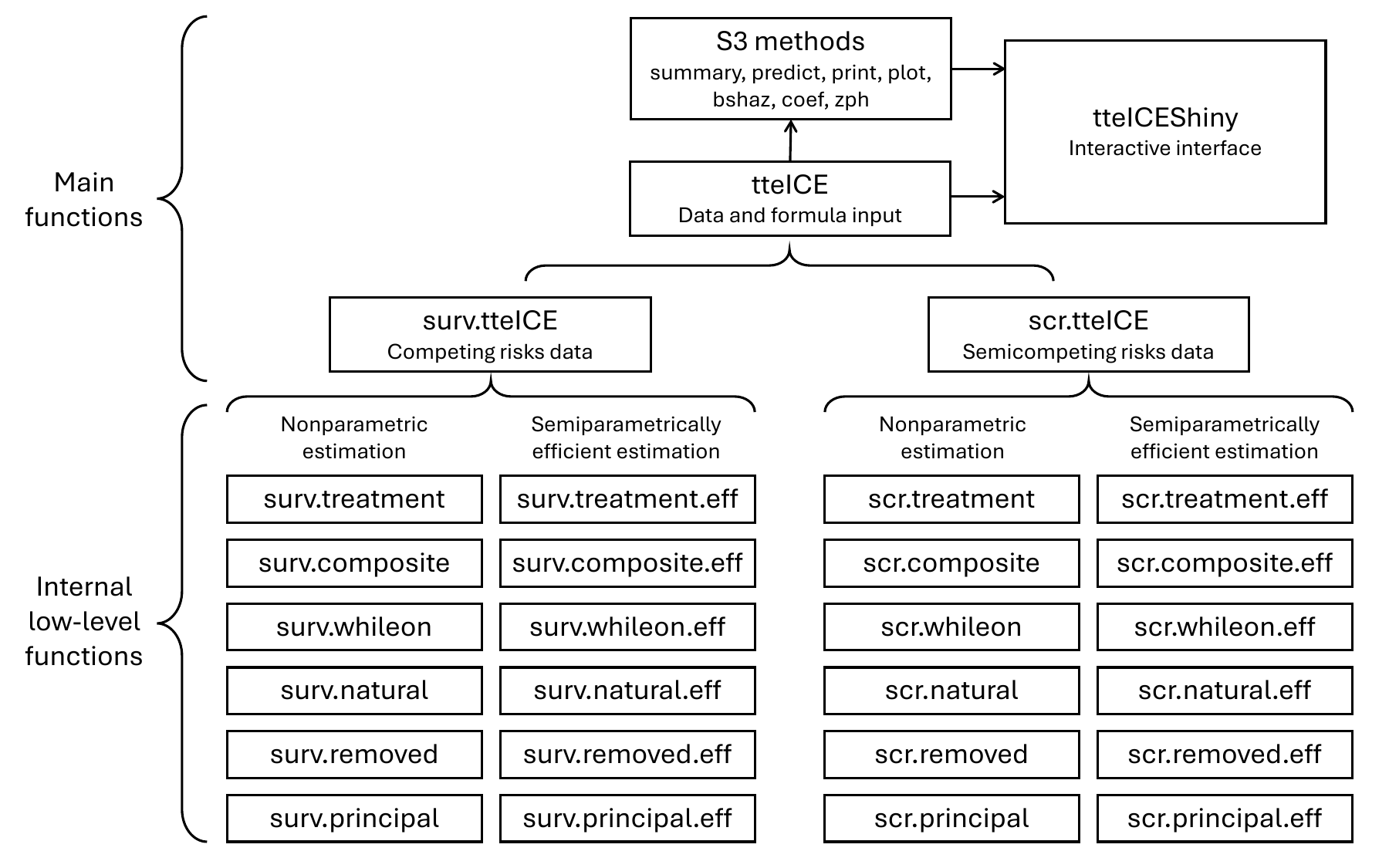}
\caption{Structure of functions in the package.} \label{pkgstr}
\end{figure}

The package is developed based on the architecture shown in Figure \ref{pkgstr}. First, we begin with nonparametric estimation for competing risks data, which is directly applicable to randomized controlled trials (Section \ref{sec:imp3}). Second, to address confounding in observational data, we implement a weighting method (Section \ref{sec:imp4}) and a semiparametrically efficient estimation approach (Section \ref{sec:imp5}). After practical justification (Section \ref{sec:imp6}), these functions are integrated into the main functions, so users can apply different strategies and estimation methods by specifying arguments in the main functions (Section \ref{sec:imp7}). S3 methods are provided for the fitted object (Section \ref{sec:imp8}). Finally, we develop a Shiny app with an interactive interface (Section \ref{sec:imp9}). Table \ref{pkgdes} lists the main functions or objects users may directly call in the \pkg{tteICE} package. Readers can directly go to Section \ref{sec:use} if they are not interested in implementation details.

\begin{table}[tbh]
\centering
\caption{Functions or objects in the package} \label{pkgdes}
\resizebox{\textwidth}{!}{
\begin{tabular}{lp{0.8\textwidth}}
\toprule
Function & Description \\
\midrule
\multicolumn{2}{l}{Main functions} \\
\code{scr.tteICE} &  Use vector and matrix data to fit CIFs for semi-competing risks time-to-event data with intercurrent events \\
\code{surv.tteICE} & Use vector and matrix data to fit CIFs for competing risks time-to-event data with intercurrent events.\\
\code{tteICE} & Use formula to fit CIFs for time-to-event data with intercurrent events, either competing risks or semi-competing risks \\
\code{tteICEShiny} & Shiny app for tteICE \\
\midrule
\multicolumn{2}{l}{Results output functions (associated methods for ``tteICE'' objects)} \\
\code{bshaz.tteICE} & Extract baseline hazards of `tteICE' objects \\
\code{coef.tteICE} & Extract coefficients of covariates in Cox models \\
\code{plot.tteICE} & Plot CIFs and treatment effects \\
\code{predict.tteICE} & Predict CIFs and treatment effects at specific time points \\
\code{print.tteICE} & Print a short summary of results from 'ttelCE' objects \\
\code{print.summary.tteICE} & Print the summary \\
\code{summary.tteICE} &  Summarize results from 'ttelCE' objects \\
\code{zph.tteICE} & Check proportional hazards assumption of the Cox models\\
\midrule
\multicolumn{2}{l}{Example data} \\
\code{bmt} & Data from Section 1.3 of \citet{klein1997survival}\\
\bottomrule
\end{tabular}
}
\end{table}

\subsection{Randomized trials: Nonparametric estimation} \label{sec:imp3}

Nonparametric estimation is suitable for completely randomized controlled trials, where there is no confounding between treatment assignment and potential times to events. No covariate adjustment is used in nonparametric estimation.

\subsubsection{Competing risks data}

When data are collected in a competing risks structure, the input data include the treatment indicator \code{A}, the time to the first event \code{Time}, and the event indicator \code{cstatus} (1 for the primary outcome event, 2 for the intercurrent event, and 0 for censoring). Suppose that the data has $n$ observations; then \code{A}, \code{Time}, and \code{cstatus} are vectors of length $n$. 

The estimands for each strategy are estimated by estimating hazard functions. By definition, the treatment policy strategy is not applicable to competing risks data because the primary outcome event may not exist for some individuals. If users apply this strategy, we proceed with the recorded values for \code{Time} and \code{cstatus}. The hazard function of the potential primary outcome event can be identified through
\begin{align*}
d\Lambda(t;w) &:= {P}(t \leq T(w) < t+dt \mid T(w) \ge t) \\
&= {P}(t \leq \tilde{T} < t+dt, \Delta^T = 1 \mid \tilde{T} \ge t, W = w),
\end{align*}
$w = 1, 0$. For the composite variable strategy, the hazard function of the composite outcome variable, the potential minimum event time $T(w) \wedge R(w)$, can be identified through
\begin{align*}
d\Lambda_{12}(t;w) &:= {P}(t \leq T(w) \wedge R(w) < t+dt \mid T(w) \wedge R(w) \ge t) \\
&= {P}(t \leq \tilde{T} \wedge \tilde{R} < t+dt, \Delta^T \vee \Delta^R = 1 \mid \tilde{T} \wedge \tilde{R} \ge t, W = w),
\end{align*}
$w=1,0$, where $\Delta^T \vee \Delta^R = \max\{\Delta^T, \Delta^R\}$. 
For other strategies, we denote the hazard specific to the primary outcome event and the hazard specific to the intercurrent event by $d\Lambda_1(t;w)$ and $d\Lambda_2(t;w)$, respectively. They are identified through
\begin{align*}
d\Lambda_1(t;w) &= {P}(t \leq T(w) < t+dt \mid T(w) \geq t, R(w) \geq t) \\
&= {P}(t \leq \tilde{T} < t+dt, \Delta^T = 1 \mid \tilde{T} \geq t, \tilde{R} \geq t, W = w), \\
d\Lambda_2(t;w) &= {P}(t \leq R(w) < t+dt \mid T(w) \geq t, R(w) \geq t) \\
&= {P}(t \leq \tilde{R} < t+dt, \Delta^R = 1 \mid \tilde{T} \geq t, \tilde{R} \geq t, W = w).
\end{align*}
respectively, where $w = 1, 0$. 
Estimation of these hazards is performed using the \fct{survfitKM} function in the \pkg{survival} package, separately for each group. The estimated cumulative hazard is extracted from \code{$cumhaz}, and the pointwise standard error is extracted from \code{$std.err}. 

\begin{itemize}
\item For the treatment policy strategy, the potential cumulative incidence function is identified as
\begin{align*}
\mu_w^{\text{tp}}(t) &= 1 - \exp\{-\Lambda(t;w)\}.
\end{align*}
The estimator is obtained by plugging in the fitted hazards. The standard error is calculated using the functional delta method. The $p$-value of the treatment effect is calculated by the logrank test for event 1 using \fct{survdiff} in the \pkg{survival} package.
\item For the composite variable strategy, the potential cumulative incidence function is identified as
\begin{align*}
\mu_w^{\text{cv}}(t) &= 1 - \exp\{-\Lambda_{12}(t;w)\}.
\end{align*}
The estimator is obtained by plugging in the fitted hazards. The standard error is calculated using the functional delta method. The $p$-value of the treatment effect is calculated by the logrank test for the composite event using \fct{survdiff} in the \pkg{survival} package.
\item For the while on treatment strategy, the potential cumulative incidence function is identified as
\begin{align*}
\mu_w^{\text{wo}}(t) &= \int_0^t \exp\{-\Lambda_{12}(s;w)\}d\Lambda_1(s;w).
\end{align*}
The estimator is obtained by plugging in the fitted hazards. The standard error is calculated using the functional delta method. The $p$-value of the treatment effect is calculated by the Gray test for event 1 using \fct{cuminc} in the \pkg{cmprsk} package.
\item For the hypothetical strategy (Scenario I, natural), the potential cumulative incidence function is identified as
\begin{align*}
\mu_w^{\text{hp,I}}(t) &= \int_0^t \exp\{-\Lambda_1(s;w) - \Lambda_2(s;0)\}d\Lambda_1(s;w).
\end{align*}
For the hypothetical strategy (Scenario II, removed), the potential cumulative incidence function is identified as
\begin{align*}
\mu_w^{\text{hp,II}}(t) &= \int_0^t \exp\{-\Lambda_1(s;w)\}d\Lambda_1(s;w) = 1 - \Lambda_1(t;w).
\end{align*}
The estimator is obtained by plugging in the fitted hazards. The standard error is calculated using the functional delta method. The $p$-value of the treatment effect is calculated by the logrank test for event 1 using \fct{survdiff} in the \pkg{survival} package. 
\item For the principal stratum strategy, under the principal ignorability assumption, the potential cumulative incidence function is identified as
\begin{align*}
\mu_w^{\text{ps}}(t) &= \frac{{P}(T(w) < t, R(w) \geq t)}{{P}(R(w) > t^*)}
= \frac{\int_0^t \exp\{-\Lambda_{12}(s;w)\} d\Lambda_1(s;w)}{1 - \int_0^{t^*} \exp\{-\Lambda_{12}(s;w)\} d\Lambda_2(s;w)}.
\end{align*}
The estimator is obtained by plugging in the fitted hazards. The standard error is calculated using the functional delta method. There is no simple form for the hypothesis test, so we leave the $p$-value as \code{NULL}.
\end{itemize}

The details of the estimates and standard errors are provided in Appendix \ref{sec:technical}.1.

\subsubsection{Semi-competing risks data}

When the data are collected in a semi-competing risks structure, the input data include the treatment indicator \code{A}, the time to the primary outcome event \code{Time} associated with its censoring indicator \code{status} (1 for event and 0 for censoring), and the time to the intercurrent event \code{Time_int} associated with its censoring indicator \code{status_int} (1 for event and 0 for censoring). They are vectors of length $n$.

\begin{itemize}
\item For the treatment policy strategy, the input of the intercurrent event (\code{Time_int} and \code{status_int}) is discarded. The arguments \code{Time} and \code{status} are passed to the algorithm for competing risks data.
\item For the composite variable strategy, while on treatment strategy, hypothetical strategy (Scenario II, removed), and principal stratum strategy, only the first occurrence of events matters. So we transform the semi-competing risks data into competing risks data using the following commands.
\begin{Code}
  Time = pmin(Time, Time_int)
  cstatus = status + 2 * status_int
  cstatus[cstatus > 2] = 2
\end{Code}
Then \code{Time} and \code{cstatus} are passed to the algorithms for competing risks data.
\item For the hypothetical strategy (Scenario I, natural), information after the intercurrent event can be used to estimate the cumulative incidence function of the primary outcome. We assume the hazard of the intermediate event is controlled at the placebo level, whereas the cause-specific hazard of the primary outcome event can vary by treatment. We define cause-specific hazards for the primary outcome event as $d\Lambda_1(t;w,m)$ assuming Markovness, where $w \in \{1,0\}$ and $m \in \{1,0\}$ is the status of the intercurrent event at time $t$. The cumulative incidence function of the primary outcome event can be written as a function of hazards,
\begin{align*}
\mu_w^{\text{hp,I}}(t) &= 1 - \exp\{-\Lambda_1(t;w,0)-\Lambda_2(t;0)\} \\
&\quad - \int_0^t\exp\{-\Lambda_1(s;w,0)-\Lambda_2(s;0)-\Lambda_1(t;w,1)+\Lambda_1(s;w,1)\}d\Lambda_2(s;0).
\end{align*}
The status of the intercurrent event acts as a time-varying covariate, possibly changing from 0 to 1. The cause-specific hazards are estimated using the Aalen--Johansen estimator. The potential cumulative incidence is then estimated by plugging in the fitted cause-specific hazards. The standard error is calculated using the functional delta method; see Appendix \ref{sec:technical}.2 for details. The $p$-value is calculated by the logrank test for event 1 using \fct{survdiff} in the \pkg{survival} package. Computation of the hypothetical strategy (Scenario I, natural) is more intensive than that of other strategies that directly use competing risks data.
\end{itemize}

\subsection{Observational studies: Inverse probability weighting} \label{sec:imp4}

Confounding needs to be addressed in observational studies. Instead of a completely randomized treatment assignment, we assume the treatment assignment is independent of potential outcomes after adjusting for a set of covariates $X$. Random censoring and positivity assumptions hold at each level of $X$.

In observational studies, inverse treatment probability weighting is a widely used method for addressing observed confounding. Let \code{X} be the baseline covariates in matrix form. We fit the propensity score by logistic regression. By default, \code{weights = rep(1, n)} if it is not supplied.
\begin{Code}
  ps = glm.fit(X, A, family = binomial(link = 'logit'), weights)
  ps = fitted(ps)
\end{Code}
The weight is calculated based on the stabilized propensity score.
\begin{Code}
  ips = A / ps * mean(A / ps) + (1 - A)/(1 - ps) * mean((1 - A)/(1 - ps))
\end{Code}
After weighting, the covariates have the same distribution in the treated and control groups. The weighted sample emulates the overall population. When estimating hazard functions, we specify the \code{weights} argument in \fct{survfit} as \code{ips}. 
However, there are two caveats to using inverse treatment probability weighting in observational studies. First, both completely random censoring and conditionally random censoring are required. Second, the model fit does not account for uncertainty in the estimated propensity score, leading to slight bias (although usually negligible) in the cumulative incidence function estimator \citep{deng2025adjusted}.

\subsection{Observational studies: Semiparametrically efficient estimation} \label{sec:imp5}

Another approach to addressing confounding in observational studies is to adjust hazard functions for covariates. Identification of potential cumulative incidence functions and treatment effects is straightforward by averaging covariate-specific quantities in the population.
Incorporating covariates into estimation uses more information, reduces estimation uncertainty, and improves robustness. Semiparametric efficient estimators are obtained by deriving the efficient influence functions (EIFs).

\subsubsection{Competing risks data}


Implementing semiparametrically efficient estimation requires fitting the propensity score and the covariate-specific hazards. The EIFs were derived in a previous methodological paper \citep[Supplementary Material]{deng2025inference}. We use \fct{coxph} with covariates \code{X} to replace the nonparametric \fct{survfitKM} to estimate hazard functions. We chose the Cox proportional hazards model as the working model for the following reasons. First, the Cox model is the most widely used semiparametric model in survival analysis and is well implemented in \proglang{R}. Prediction with the Cox model can be performed using existing \proglang{R} functions. Second, the Cox model offers an easier interpretation than other survival models, as its coefficients are interpreted as log hazard ratios. Third, the Cox model yields more computationally reliable results than the additive hazards model, as the baseline hazard estimated by the Cox model is always non-negative, whereas that estimated by the additive hazards model can be negative when the sample size is limited. The baseline cumulative hazards are extracted from \code{$cumhaz} and the coefficients of covariates are extracted from \code{$coefficients}.

Let $\widehat{P}(W=w \mid X=x)$ be the estimated stabilized propensity score. Let $\Lambda_1(t|w,x)$ be the hazard for the primary outcome event, which is estimated by $\widehat\Lambda_1(t|w,x)$. Let $\Lambda_2(t|w,x)$ be the hazard for the intercurrent event, which is estimated by $\widehat\Lambda_2(t|w,x)$. Let $\Lambda_0(t|w,x)$ be the hazard of censoring, which is estimated by $\Lambda_0(t|w,x)$. We code event 1 as the primary outcome event and event 2 as the intercurrent event. The martingale process of the $j$th event ($j=1,2$) is $M_j(t|w,x) = \int_0^t \{dN_j(s;w) - Y(s;w)d\Lambda_j(s|w,x)\}$, where $N_j(t;w)$ is the counting process of the $j$th event and $Y(t;w)$ is the at-risk process shared for both events in treatment group $w = 1, 0$. Let $\widehat{M}_j(t|w,x)$ be the fitted martingale process. Let $\widehat\Lambda_{12}(t|w,x) = \widehat\Lambda_{1}(t|w,x) + \widehat\Lambda_{2}(t|w,x)$ and $\widehat{M}_{12}(s|w,x) = \widehat{M}_{1}(s|w,x) + \widehat{M}_{2}(s|w,x)$. 

\begin{itemize}
\item For the treatment policy strategy, the estimated cumulative incidence function $\widehat{\mu}_w^{\text{tp}}(t)$ associated with treatment condition $w$ is the sample average of the non-centered EIF
\begin{align*}
 \frac{I(W=w)}{\widehat{{P}}(W=w\mid X)} e^{-\widehat\Lambda_1(t|w,X)} \int_0^t \frac{d\widehat{M}_1(s|w,X)}{e^{-\widehat\Lambda_{12}(s|w,X)-\widehat\Lambda_0(s|w,X)}} + 1 - e^{-\widehat\Lambda_1(t|w,X)}.
\end{align*}
\item For the composite variable strategy, the estimated cumulative incidence function $\widehat{\mu}_w^{\text{cv}}(t)$ associated with treatment condition $w$ is the sample average of the non-centered EIF
\begin{align*}
\frac{I(W=w)}{\widehat{{P}}(W=w\mid X)} e^{-\widehat\Lambda_{12}(t|w,X)} \int_0^t \frac{d\widehat{M}_{12}(s|w,X)}{e^{-\widehat\Lambda_{12}(s|w,X)-\widehat\Lambda_0(s|w,X)}} + 1 - e^{-\widehat\Lambda_{12}(t|w,X)}.
\end{align*}
For the while on treatment strategy, we denote the regression-based estimate of the cumulative incidence function as $\widehat\mu^{\text{wo}}(t|w,x)=\int_0^t e^{-\widehat\Lambda_{12}(s|w,x)}d\widehat\Lambda_1(s|w,x)$. The estimated cumulative incidence function $\widehat{\mu}_w^{\text{wo}}(t)$ associated with treatment condition $w$ is the sample average of the non-centered EIF
\begin{align*}
&\frac{I(W=w)}{\widehat{{P}}(W=w\mid X)} \int_0^t \left\{1-\frac{\widehat\mu^{\text{wo}}(t|w,X)-\widehat\mu^{\text{wo}}(s|w,X)}{e^{-\widehat\Lambda_{12}(s|w,X)}}\right\}\frac{d\widehat{M}_1(s|w,X)}{e^{-\widehat\Lambda_0(s|w,X)}} \\
& - \frac{I(W=w)}{\widehat{{P}}(W=w\mid X)} \int_0^t \frac{\widehat\mu^{\text{wo}}(t|w,X)-\widehat\mu^{\text{wo}}(s|w,X)}{e^{-\widehat\Lambda_{12}(s|w,X)}}\frac{d\widehat{M}_2(s|w,X)}{e^{-\widehat\Lambda_0(s|w,X)}} + \widehat\mu^{\text{wo}}(t|w,X).
\end{align*}
\item For the hypothetical strategy (Scenario I, natural), we denote the regression-based estimate of the cumulative incidence function as $\widehat\mu^{\text{hp,I}}(t|w,x)=\int_0^t e^{-\widehat\Lambda_{1}(s|w,x)-\widehat\Lambda_{2}(s|0,x)}d\widehat\Lambda_1(s|w,x)$.
The estimated cumulative incidence function $\widehat{\mu}_w^{\text{hp,I}}(t)$ associated with treatment condition $w$ is the sample average of the non-centered EIF
\begin{align*}
&\frac{I(W=w)}{\widehat{{P}}(W=w\mid X)} \int_0^t \left\{\frac{e^{-\widehat\Lambda_2(s|0,X)}}{e^{-\widehat\Lambda_2(s|w,X)}}-\frac{\widehat\mu^{\text{hp,I}}(t|w,X)-\widehat\mu^{\text{hp,I}}(s|w,X)}{e^{-\widehat\Lambda_1(s|w,X)-\widehat\Lambda_2(s|w,X)}}\right\}\frac{d\widehat{M}_1(s|w,X)}{e^{-\widehat\Lambda_0(s|w,X)}} \\
& - \frac{I(W=0)}{\widehat{{P}}(W=0\mid X)} \int_0^t \frac{\widehat\mu^{\text{hp,I}}(t|w,X)-\widehat\mu^{\text{hp,I}}(s|w,X)}{e^{-\widehat\Lambda_1(s|0,X)-\widehat\Lambda_2(s|0,X)}}\frac{d\widehat{M}_2(s|0,X)}{e^{-\widehat\Lambda_0(s|0,X)}} + \widehat\mu^{\text{hp,I}}(t|w,X).
\end{align*}
For the hypothetical strategy (Scenario II, removed), the estimated cumulative incidence function $\widehat{\mu}_w^{\text{hp,II}}(t)$ associated with treatment condition $w$ is the sample average of the non-centered EIF
\begin{align*}
\frac{I(W=w)}{\widehat{{P}}(W=w\mid X)} e^{-\widehat\Lambda_{1}(t|w,X)} \int_0^t \frac{d\widehat{M}_{1}(s|w,X)}{e^{-\widehat\Lambda_{12}(s|w,X)-\widehat\Lambda_0(s|w,X)}} + 1 - e^{-\widehat\Lambda_{1}(t|w,X)}.
\end{align*}
\item For the principal stratum strategy, typically, identification of the principal stratum estimand requires an additional monotonicity assumption; otherwise, the proportion of the stratum $\{R(1)=R(0)=\infty\}$ is not identifiable. To reduce the assumptions involved and computational complexity, we consider a pseudo-estimand ${P}(T(1)<t \mid R(1)=\infty) - {P}(T(0)<t \mid R(0)=\infty)$ instead. This estimand may not have a meaningful causal interpretation unless there is no systematic difference between the populations with $\{R(1)=\infty\}$ and those with $\{R(0)=\infty\}$.
The estimated cumulative incidence function associated with treatment condition $w$ is the sample average of the non-centered EIF
\begin{align*}
\frac{\widehat{\mathbb{IF}}\{\widehat\mu_w^{\text{wo}}(t)\}+\widehat\mu_w^{\text{wo}}(t)}{1-\widehat\mu_w^{\text{cv}}(t^*)+\widehat\mu_w^{\text{wo}}(t^*)} + \frac{\widehat\mu_w^{\text{wo}}(t) [\widehat{\mathbb{IF}}\{\widehat\mu_w^{\text{cv}}(t^*)-\widehat\mu_w^{\text{wo}}(t^*)\}]}{\{1-\widehat\mu_w^{\text{cv}}(t^*)+\widehat\mu_w^{\text{wo}}(t^*)\}^2},
\end{align*}
where $\mathbb{IF}(\cdot)$ is the non-centered EIF operator, which is an intermediate step in the estimation under the composite variable and the while on treatment strategies.
\end{itemize}
For all five strategies, standard errors are calculated based on the standard deviations of the estimated non-centered EIFs; see Appendix \ref{sec:technical}.2 for details. Hypothesis tests are based on the restricted mean survival time lost by the end of the study. The estimators based on EIFs are asymptotically efficient: they have the smallest asymptotic variances among all regular and asymptotically linear estimators for the target estimands when the models are correctly specified. Another appealing property is multiple robustness. There are three sets of models in general: (1) propensity score and censoring hazard, (2) primary outcome event hazard, and (3) intercurrent event hazard. Whenever two of the three sets of models are correctly specified, the estimators are consistent. Under partial model misspecification, the standard error based on EIFs is not accurate; bootstrapping standard error should be used instead.

\subsubsection{Semi-competing risks data}

Following the same roadmap, we reduce the analysis of semi-competing risks data to competing risks except for the hypothetical strategy (Scenario I, natural).

\begin{itemize}
\item For the treatment policy strategy, the time to the primary outcome event \code{Time}, along with its censoring indicator \code{status}, will be passed to the algorithm for competing risks data. Information about the intercurrent event will be discarded.
\item For the composite variable strategy, while on treatment strategy, hypothetical strategy (Scenario II, removed), and principal stratum strategy, only the first occurrence of events matters. So we transform the semi-competing risks data into competing risks data as described in Section \ref{sec:imp3}.
\item For the hypothetical strategy (Scenario I, natural), we need to estimate the cause-specific hazards of the primary outcome event, accounting for the status of the intercurrent event. We assume Markovness; that is, the cause-specific hazard of the primary outcome event depends only on the status of ICEs (whether they occurred), not on their timing. Let $\widehat\Lambda_1(t|w,x)$ be the estimated cause-specific hazard of the primary outcome event with no history of ICEs, $\widehat\Lambda_2(t|w,x)$ be the estimated hazard of ICEs, and $\widehat\Lambda_3(t|w,x)$ be the estimated cause-specific hazard of the primary outcome event with a history of ICEs. Let $\widehat{M}_1(t|w,x)$, $\widehat{M}_2(t|w,x)$, and $\widehat{M}_3(t|w,x)$ be the corresponding estimated martingale processes. The estimated hazard of the first occurrence of events $\widehat\Lambda_{12}(t|w,x) = \widehat\Lambda_{1}(t|w,x)+\widehat\Lambda_{2}(t|w,x)$. The EIF is more complex in this setting, but can still be expressed in a function of observed data, including hazards and martingales \citep{breum2024estimation}.
The estimated cumulative incidence function associated with treatment condition $w$ is the sample average of the non-centered EIF
\begin{align*}
&\int_0^t \exp\{-\widehat\Lambda_1(s|w,X)\} \bigg\{\frac{I(W=w)}{\widehat{P}(W=w\mid X)} \frac{d\widehat{M}_1(s|w,X)}{e^{-\widehat\Lambda_{12}(s|w,X)-\widehat\Lambda_0(s|w,X)}} \\
&\quad - \frac{I(W=w)}{\widehat{P}(W=w\mid X)} \int_0^s \frac{d\widehat{M}_1(u|w,X) d\widehat\Lambda_1(s|w,X)}{e^{-\widehat\Lambda_{12}(u|w,X)-\widehat\Lambda_0(u|w,X)}} \\
&\quad - \frac{I(W=0)}{\widehat{P}(A=0\mid X)} \int_0^s \frac{d\widehat{M}_2(u|0,X) d\widehat\Lambda_1(s|w,X)}{e^{-\widehat\Lambda_{12}(u|0,X)-\widehat\Lambda_0(u|0,X)}} \bigg\} \\
&\quad + \int_0^t\int_0^s \exp\{-\widehat\Lambda_1(r|w,X)-\widehat\Lambda_2(r|0,X)-\widehat\Lambda_3(s|w,X)+\widehat\Lambda_3(r|w,X)\} \\
&\quad \bigg\{\frac{I(W=0)}{\widehat{P}(W=0\mid X)} \frac{d\widehat{M}_2(r|0,X) d\widehat\Lambda_3(s|w,X)}{e^{-\widehat\Lambda_{12}(r|0,X)-\widehat\Lambda_0(r|0,X)}} \\
&\qquad - \frac{I(W=w)}{\widehat{P}(W=w\mid X)} \int_0^r \frac{d\widehat{M}_1(u|w,X) d\widehat\Lambda_2(r|0,X)d\widehat\Lambda_3(s|w,X)}{e^{-\widehat\Lambda_{12}(u|w,X)-\widehat\Lambda_0(u|w,X)}}  \\
&\qquad - \frac{I(W=0)}{\widehat{P}(W=0\mid X)} \int_0^r \frac{d\widehat{M}_2(u|0,X) d\widehat\Lambda_2(r|0,X)d\widehat\Lambda_3(s|0,X)}{e^{-\widehat\Lambda_{12}(u|0,X)-\widehat\Lambda_0(u|0,X)}}  \\
&\qquad + \frac{I(W=w)}{\widehat{P}(W=w\mid X)} \frac{d\widehat{M}_3(s|w,X) d\widehat\Lambda_2(s|0,X)}{\int_0^s e^{-\widehat\Lambda_{12}(u|w,X)-\widehat\Lambda_3(s|w,X)-\widehat\Lambda_0(s|w,X)+\widehat\Lambda_3(u|w,X)}d\widehat\Lambda_2(u|w,X)} \\
&\qquad - \frac{I(W=w)}{\widehat{P}(W=w\mid X)} \int_0^s\frac{d\widehat{M}_3(u|w,X) d\widehat\Lambda_2(r|0,X)d\widehat\Lambda_3(s|w,X)}{\int_0^u e^{-\widehat\Lambda_{12}(v|w,X)-\widehat\Lambda_3(v|w,X)-\widehat\Lambda_0(u|w,X)+\widehat\Lambda_3(u|w,X)}d\widehat\Lambda_2(v|w,X)} \bigg\} \\
&\quad + \int_0^t e^{-\widehat\Lambda_1(s|w,X)-\widehat\Lambda_2(s|0,X)}d\widehat\Lambda_1(s|w,X) \\
&\quad + \int_0^t\int_0^s e^{-\widehat\Lambda_1(r|w,X)-\widehat\Lambda_2(r|0,X)-\widehat\Lambda_3(s|w,X)+\widehat\Lambda_3(r|w,X)}d\widehat\Lambda_2(r|0,X)d\widehat\Lambda_3(s|w,X).
\end{align*}
We use the \fct{tmerge} function in the \pkg{survival} package to transform the time to the primary outcome event into intervals split by the first occurrence of the ICE. Then, the estimation of $\Lambda_1(t|w,x)$ and $\Lambda_3(t|w,x)$ is achieved by a single Cox model with a time-varying covariate $I(\tilde{R}<t, \Delta^R=1)$. The estimator of the potential cumulative incidence function under the hypothetical strategy (Scenario I, natural) is substantially more complex and computationally more intensive than those for other strategies.
\end{itemize}

\section{Practical usage of the package tteICE} \label{sec:use}

The package is officially available from the Comprehensive R Archive Network (CRAN) at \url{https://cran.r-project.org/package=tteICE}. The developed version is accessible on GitHub at \url{https://mephas.github.io/tteICE/}. The primary functions for statistical analysis include \fct{surv.tteICE} for competing risks data, \fct{scr.tteICE} for semi-competing risks data, and \fct{tteICE} for both. These functions return a `\code{tteICE}' object containing estimated cumulative incidence functions, treatment effects, and corresponding standard errors. Important S3 methods, including \code{summary}, \code{print}, \code{plot}, and \code{predict}, are implemented to summarize a list of main estimation results, present a formatted summary of those results, plot the estimated cumulative incidence functions and treatment effects, and those at specified time points, respectively. In the following, we outline the preparations required before using the package and the estimation methods implemented within it.

\subsection{Preparations before using the tteICE package} \label{sec:imp6}

For investigators seeking to apply these methods to observational studies, we recommend first emulating a target trial to clarify the causal question and minimize confounding \citep{hernan2016using}. This process involves explicitly defining the key components of a hypothetical randomized trial---including eligibility criteria, treatment strategies, assignment procedures, outcomes, follow-up periods, and causal contrasts---using the observational dataset. Once this structure is in place and the data have been appropriately structured to mimic the protocol of the target trial (e.g., by defining treatment cohorts at a baseline time zero and adjusting for confounding through matching) \citep{ho2011matchit, sekhon2011multivariate}, the functions within this package can be applied. The \pkg{tteICE} package includes two methods to address confounding: inverse treatment probability weighting and semiparametrically efficient estimation, as described in Sections \ref{sec:imp4} and \ref{sec:imp5}.

Furthermore, the validity of all subsequent estimates is contingent upon the appropriate handling of missing data. Before applying any functions of this package, investigators must address missingness in key variables, including covariates used for adjustment, treatment assignment indicators, and outcome variables (follow-up times and event statuses). It is critical to evaluate the mechanism of missingness (e.g., missing completely at random, missing at random) to inform the choice of handling method, as analyses conducted on a dataset with unaddressed missingness may yield biased and misleading conclusions. The functions in \pkg{tteICE} package only analyze complete cases and simply discard observations with missingness. We recommend employing robust methods, such as multiple imputation, to handle missing covariate data using the \pkg{mice} package \citep{van2011mice}, as this preserves the sample size and reduces potential bias from complete-case analysis. The multiply imputed datasets can then be analyzed separately using the functions in this package, with results pooled according to Rubin's rules \citep{rubin2018multiple}.

\subsection{Main functions} \label{sec:imp7}

The function \fct{surv.tteICE} analyzes competing risks data, and the function \fct{scr.tteICE} analyzes semi-competing risks data. Users can specify the strategy and estimation methods. The \code{strategy} argument indicates the strategy used, and the \code{method} argument indicates the estimation methods. If \code{method = `np'} (default), then nonparametric estimation is used. If \code{method = `ipw'}, then inverse treatment probability weighting is used. If \code{method = `eff'}, then semiparametrically efficient estimation is used. The \code{cov1} argument represents covariates.
\begin{Code}
  fit <- surv.tteICE(A, Time, cstatus, strategy = 'composite',  
           cov1 = NULL, method = 'np', weights = NULL, subset = NULL, 
           na.rm = FALSE, nboot = 0, seed = NULL)
  fit <- scr.tteICE(A, Time, status, Time_int, status_int, 
           strategy = 'composite', cov1 = NULL, method = 'np', 
           weights = NULL, subset = NULL, na.rm = FALSE, 
           nboot = 0, seed = NULL) 
\end{Code}
The \code{subset} argument can be specified to select a subset of the sample for analysis. The covariates \code{cov1} will not be used for nonparametric estimation, and the \code{weights} will not be used for semiparametrically efficient estimation, even if they are supplied. The \code{nboot} argument determines how the standard error is calculated. If \code{nboot = 0}, then asymptotic formulas will be used. If \code{nboot > 1}, then bootstrapping will be used by calling the \fct{surv.boot} function. The \code{seed} sets the random seed. If \code{seed = NULL}, a default seed will be used. For computational efficiency, the asymptotic standard error is recommended.

Based on the two functions above, a unified function \fct{tteICE} was developed that takes the formula interface as input. The outcome of the formula relies on the \code{Surv()} function defined in the R package \pkg{survival}. According to the \code{Surv()} function, for competing risk outcomes, the \code{cstatus} should be a factor. The primary outcome event is coded as 1, intercurrent events as 2, and censoring as 0. For semi-competing risk data, in addition to the primary outcomes, the intercurrent event time and status should be defined in the argument \code{add.scr} as follows.
\begin{Code}
  fit <- tteICE(Surv(Time, cstatus) ~ A, 
           data, strategy = 'composite',  
           method = 'np', weights = NULL, subset = NULL, 
           na.rm = FALSE, nboot = 0, seed = NULL)
  fit <- tteICE(Surv(Time, status) ~ A, 
           add.scr = ~ Surv(Time_int, status_int), data, 
           strategy = 'composite', method = 'np', weights = NULL, 
           subset = NULL, na.rm = FALSE, nboot = 0, seed = NULL) 
\end{Code}
If baseline covariates need to be controlled, the first formula interface can be correspondingly written as \code{Surv(Time, cstatus) ~ A | X1 + X2} or \code{Surv(Time, status) ~ A | X1 + X2}, where \code{X1} and \code{X2} are the names of two covariates. 

The program will automatically check whether \code{A} is binary. If not, the function terminates, and an error message is displayed. If \code{strategy} is not one of the six possible choices, a warning message will be displayed, and \code{`composite'} will be used as the default. If \code{method} is not any one of the three possible choices, a warning message will be displayed, and \code{`np'} will be used as the default. If \code{na.rm = TRUE}, the program will check whether there are missing data and delete the cases with missing data.
The output is an object of class `\code{tteICE}', for which standard S3 methods are available. The returned object contains time points, estimated cumulative incidence functions, treatment effects, standard errors, and $p$-values (if any). 

\subsection{S3 methods for tteICE objects} \label{sec:imp8}

For a \code{tteICE} object fitted by the above functions, the S3 method \code{summary} is available to display summaries of the fitted object, and \code{print} is available to display treatment effects at quarterly follow-up times.
\begin{Code}
  summary(fit)
  print(fit)
\end{Code}

We created two internal functions to graphically present the estimated results. The function \fct{plot\_inc} is used to plot the estimated cumulative incidence functions.
\begin{Code}
  tteICE:::plot_inc(fit, decrease = FALSE, conf.int = .95, xlab = 'Time', 
                    plot.configs = list(...), ...)
\end{Code}
Here, \code{fit} is a `\code{tteICE}' object, \code{decrease} indicates whether to plot the cumulative incidence function (increasing with time) or survival function (decreasing with time). If \code{conf.int} is not \code{NULL}, then the confidence interval at that level is displayed. The confidence interval can be obtained based on asymptotic formulas (which we use to calculate the standard error) or bootstrapping, depending on the output of \code{fit}. Other standard plot arguments (such as \code{legend}, \code{cex}, \code{ylim}, etc.) can be customized by specifying them in \code{plot.configs = list(...)}.

Another function \fct{plot\_ate} is used to plot the estimated treatment effects.
\begin{Code}
  tteICE:::plot_ate(fit, decrease = FALSE, conf.int = .95, xlab = 'Time',
                    plot.configs = list(...), ...)
\end{Code}
In this function, if \code{decrease = FALSE}, it shows the difference in potential cumulative incidence functions under treated and control. If \code{decrease = TRUE}, the plot shows the negative difference in potential cumulative incidence functions under treated and control. If \code{con.int} is not \code{NULL}, then the confidence interval at that level is displayed. The confidence interval can be obtained based on asymptotic formulas (which we use to calculate the standard error) or bootstrapping. If \code{nboot} $\leq$ 1, asymptotic formulas will be used. If \code{nboot} $>$ 1, then bootstrapping will be used. For computational efficiency, the asymptotic confidence interval is recommended. Similarly, other standard plot arguments, such as \code{legend}, \code{cex}, \code{ylim}, etc., can be customized by specifying them in \code{plot.configs = list(...)}.

The corresponding S3 method \fct{plot} can be called as follows. 
\begin{Code}
  plot(fit, type = c("ate", "inc")[1], ...)
\end{Code}
The display of the estimated treatment effects or cumulative incidence functions is controlled by the \code{type} argument.

The S3 method \code{predict} is implemented to estimate or predict the potential cumulative incidence functions and treatment effects at specific time points. 
\begin{Code}
  predict(fit, timeset = NULL)
\end{Code}
When no time point is specified, such as \code{timeset = NULL}, predictions are performed at the quartiles of the maximum follow-up time.

The model fit can be assessed when semiparametrically efficient estimation is used. The function \fct{coef} extracts the covariate coefficients from the Cox models fitted in each treatment group and returns them as a list. The function \fct{basehaz} extracts the baseline cumulative hazards. The function \fct{zph} tests the proportional hazards assumption based on the Schoenfeld residuals stored as a table, including $p$-values for each covariate separately and the $p$-value for a global test \citep{Grambsch1994proportional}.
\begin{Code}
  coef(fit)
  bshaz(fit)
  zph(fit)
\end{Code}
Due to the multiple robustness of the semiparametrically efficient estimator, the resulting estimates remain consistent even if the proportional hazards assumption fails. However, when there is evidence of a violation of the proportional hazards assumption, users should exercise greater caution when interpreting the coefficients, as they are weighted averages of the time-varying log hazard ratio. Although the asymptotic standard error is no longer accurate, bootstrapping standard error remains reliable.

\subsection{Shiny app} \label{sec:imp9}

In our package, we provide an interactive app built upon the \pkg{shiny} package \citep{shiny}. 
\begin{Code}
  tteICEShiny()
\end{Code}

\begin{figure}
    \centering
    \includegraphics[width=0.95\textwidth]{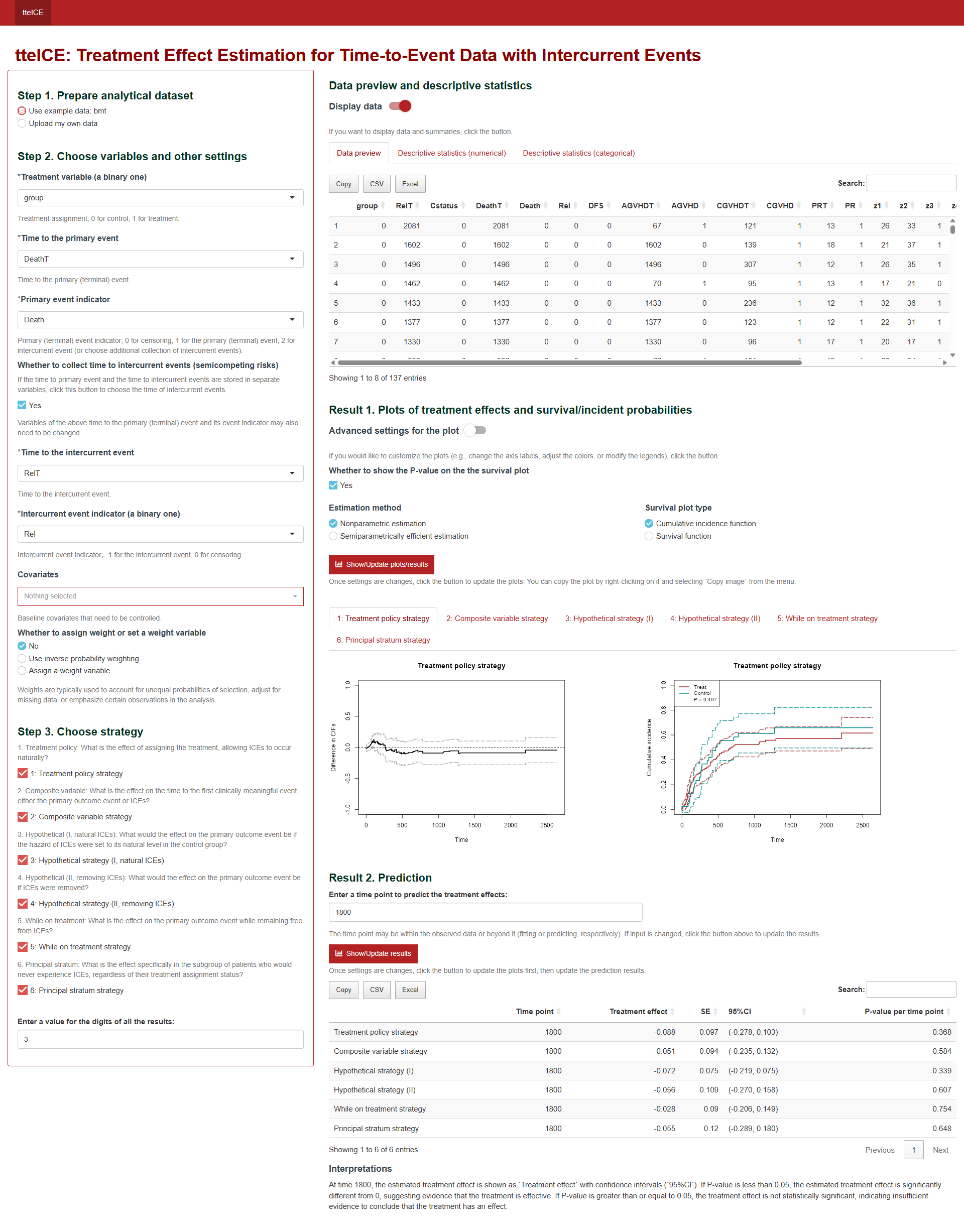}
    \caption{The interface of the \pkg{tteICE} Shiny app.} \label{fig:main1}
\end{figure}

Figure \ref{fig:main1} shows the interface of the Shiny app. The left panel allows users to upload datasets and specify variables and other parameters, while the right panel displays the generated results. The interactive interface provides step-by-step instructions for using the main functions of our package, including uploading data, selecting options, and displaying analysis results. 

In the top-right panel, a button is provided to preview the datasets and generate summary statistics. By default, users can select the \code{Time} and \code{cstatus} from the data in the competing risks structure. If the structure of semi-competing risks data is desired, users can check the ``semi-competing risks'' option and then select two additional variables. After specifying the input data, users can choose whether to use weighting methods and select the appropriate strategy. 

\begin{figure}
    \centering
    \includegraphics[width=0.95\textwidth]{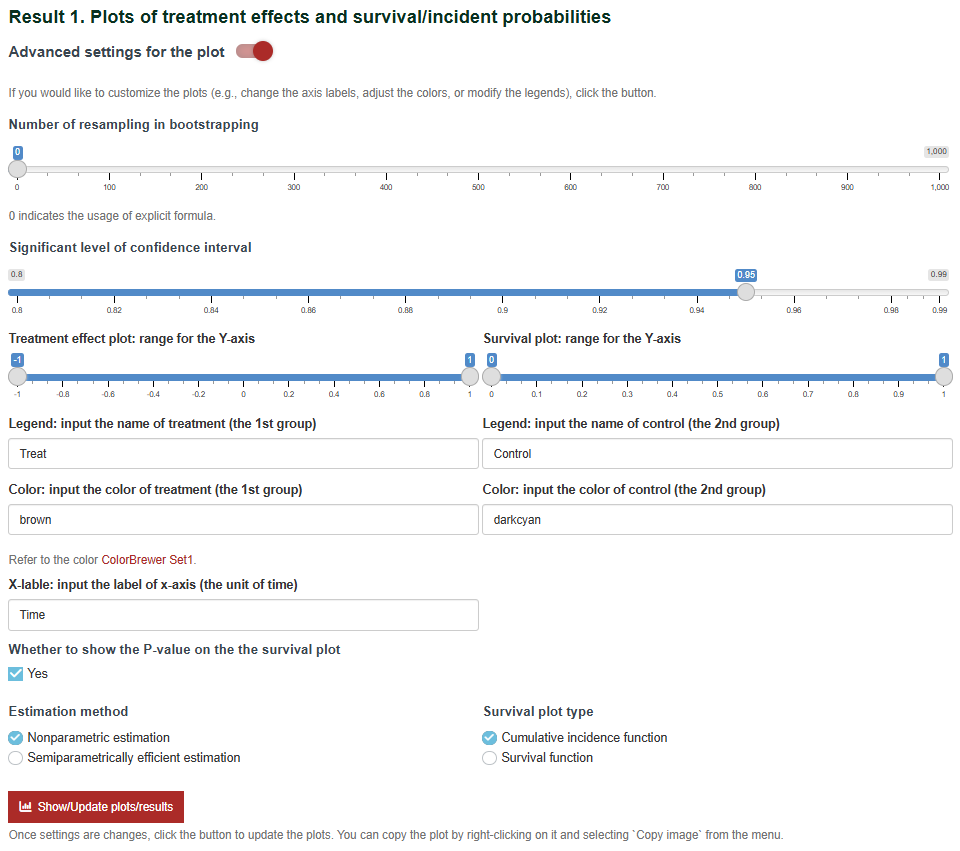}
    \caption{The advanced setting for the plots in the \pkg{tteICE} Shiny app.} \label{fig:main2}
\end{figure}

\begin{figure}
    \centering
    \includegraphics[width=0.95\textwidth]{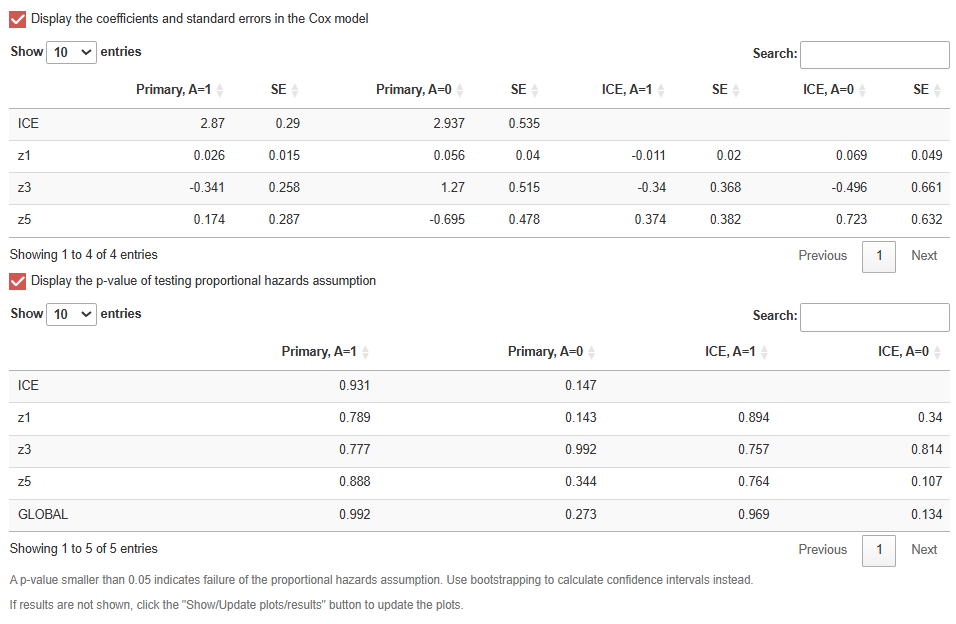}
    \caption{The model diagnosis for semiparametrically efficient estimation in the \pkg{tteICE} Shiny app.} \label{fig:main3}
\end{figure}

Advanced options for plots are offered, including the confidence-interval calculation method, the estimation method, the significance level, the $y$-axis range, and the legend names and corresponding colors (Figure \ref{fig:main2}). By clicking the ``Show/Update plots'' button, the estimated treatment effect curve and cumulative incidence curve will be plotted. If semiparametrically efficient estimation is used, an additional model diagnosis is valid after the main analysis. The coefficients in the working Cox models can be displayed, associated with standard errors and $p$-values of the proportional hazards test (Figure \ref{fig:main3}).
Once the plots are generated, the predicted treatment effect (with confidence interval) and $p$-value at any specified time point can be calculated.


\section{Application to bone marrow transplantation (BMT) data} \label{sec:illust}

In the following, we illustrate the implementation of this package using a real-world dataset. Simulation studies were also conducted to assess the package's validity and accuracy. The detailed simulation studies and results are presented in Appendix \ref{sec:simulation}.

This dataset was introduced in Section 1.3 of the book by \citet{klein1997survival}, which includes 137 individuals and 22 variables. These patients were prepared for transplantation with a radiation-free conditioning regimen. The preparative regimen used in this study of allogeneic marrow transplants for patients with acute myeloid leukemia (AML) and acute lymphoblastic leukemia (ALL) consisted of 16 mg/kg oral Busulfan (BU) and 120 mg/kg intravenous cyclophosphamide (Cy). A total of 137 patients 
(99 AML, 38 ALL) were treated at one of four hospitals: 76 at the Ohio State University Hospitals in Columbus, 21 at Hahnemann University in Philadelphia, 23 at St. Vincent's Hospital in Sydney, Australia, and 17 at Alfred Hospital in Melbourne. The study comprises transplants performed at these institutions from March 1, 1984, to June 30, 1989. The maximum follow-up was 7 years. There were 42 patients who relapsed and 41 who died while in remission. Several potential risk factors were measured at the time of transplantation.

The dataset \code{bmt} is embedded in the \pkg{tteICE} package. We first install the \pkg{tteICE} package from either CRAN,
\begin{CodeChunk}
\begin{CodeInput}
R> install.packages("tteICE")
\end{CodeInput}
\end{CodeChunk}
or GitHub,
\begin{CodeChunk}
\begin{CodeInput}
R> install.packages("pak")
R> pak::pak("mephas/tteICE")
\end{CodeInput}
\end{CodeChunk}

The primary outcome event is death, with follow-up time \code{t1} and indicator \code{d1} (1 for event, 0 for censoring). The intercurrent event is relapse, with follow-up time \code{t2} and indicator \code{d2} (1 for event, 0 for censoring). In addition, a disease-free survival indicator is recorded in \code{d3}, calculated as the maximum of \code{d1} and \code{d2}. For illustrative purposes, we select three covariates: patient age in years, patient sex (1 for male, 0 for female), and patient Cytomegalovirus status (1 for positive, 0 for negative). We consider AML as the treatment and ALL as the control.

\begin{CodeChunk}
\begin{CodeInput}
R> library(tteICE)
R> data(bmt)
R> A = as.numeric(bmt$group > 1)
R> X = as.matrix(bmt[, c('z1', 'z3', 'z5')])
R> bmt = transform(bmt, d4 = d2 + d3)
\end{CodeInput}
\end{CodeChunk}

There are two routes to analyze the treatment effect on death by adjusting for relapse. The first is to use competing risks data, where the data after disease-free survival are not used. The second is to use semi-competing risks data, in which both the times to relapse and to death are used.

\subsection{Analysis in the competing risks setting}

The competing risks data include the disease-free survival time \code{t2} and event indicator \code{d4} (1 for death, 2 for relapse, and 0 for censoring). First, we use nonparametric estimation. The covariates \code{X} will not be used, even if supplied in the function. For illustration, we consider all available strategies for competing risks data, including the composite variable strategy, the while on treatment strategy, the hypothetical strategy (Scenario I, natural; Scenario II, removed), and the principal stratum strategy.

\begin{CodeChunk}
\begin{CodeInput}
R> slist = c("composite", "whileon", "natural", "removed", "principal")
R> for (st in slist){
R>   fit = surv.tteICE(A, bmt$t2, bmt$d4, st)
R>   plot(fit, type = "inc", plot.configs = list(legend = c('AML', 'ALL')))
R>   plot(fit, type = "ate")
R> }
\end{CodeInput}
\end{CodeChunk}

\begin{figure}
    \centering
    \includegraphics[width=0.95\textwidth]{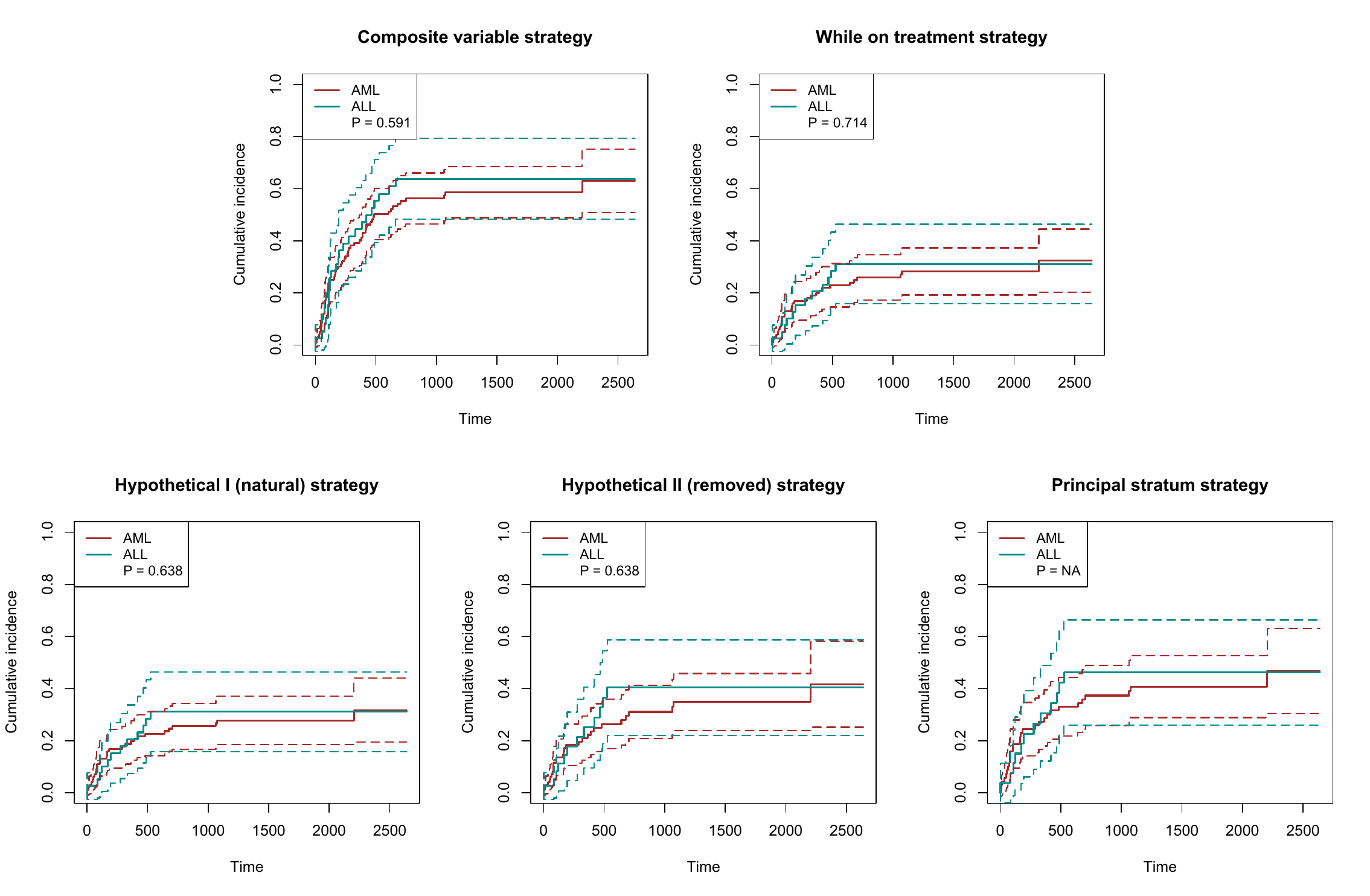}
    \caption{Analysis 1 for BMT data: cumulative incidence functions. Competing risks data, nonparametric estimation.} \label{fig:bmt1}
\end{figure}

\begin{figure}
    \centering
    \includegraphics[width=0.95\textwidth]{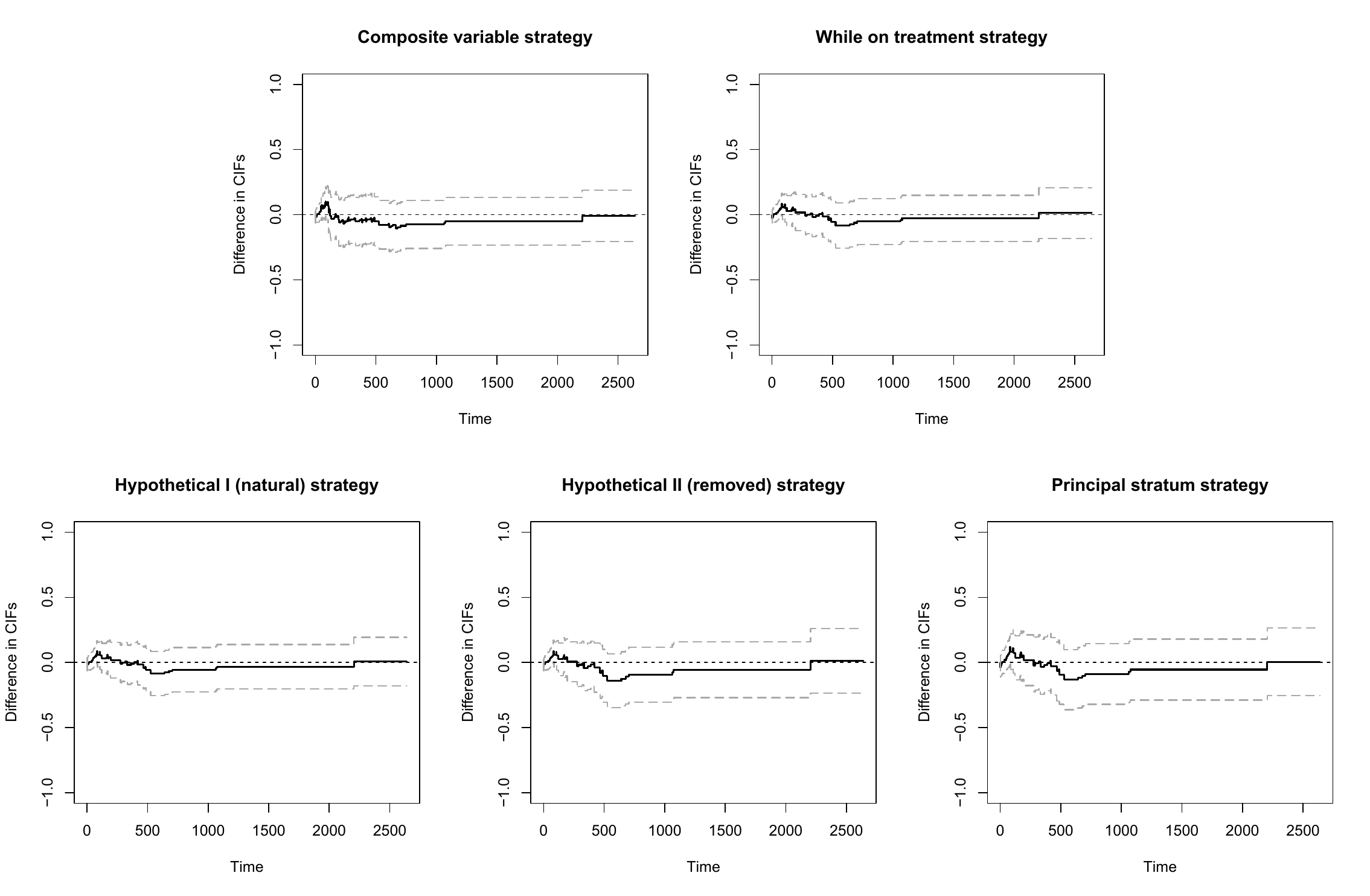}
    \caption{Analysis 1 for BMT data: treatment effects. Competing risks data, nonparametric estimation.} \label{fig:bmt1_ate}
\end{figure}

Figure \ref{fig:bmt1} shows the estimated cumulative incidence functions under AML and ALL using nonparametric estimation. $p$-values are available for all strategies except the principal stratum strategy. It is obvious that the estimated cumulative incidence functions are larger under the composite variable strategy than under the while on treatment and hypothetical strategies, because the composite outcome includes more events. In addition, the estimated cumulative incidence functions are higher under the hypothetical strategy (Scenario II, removed) than under the hypothetical strategy (Scenario I, natural), because the at-risk sets are larger when intercurrent events are removed. Figure \ref{fig:bmt1_ate} shows the estimated treatment effect under AML and ALL using nonparametric estimation. The treatment effect is the difference in cumulative incidence functions between AML and ALL. In these figures, the 95\% confidence intervals are obtained by asymptotic formulas.


Next, we use semiparametrically efficient estimation. To illustrate the commands in the plot function, we plot the survival functions rather than the cumulative incidence functions. The $p$-values are available for all strategies.

\begin{CodeChunk}
\begin{CodeInput}
R> slist = c("composite", "whileon", "natural", "removed", "principal")
R> for (st in slist){
R>   fit = surv.tteICE(A, bmt$t2, bmt$d4, st, X, method = "eff")
R>   plot(fit, type = "inc", decrease = TRUE, 
R>        plot.configs = list(legend = c('AML', 'ALL')))
R>   plot(fit, type = "ate", decrease = TRUE)
R> }
\end{CodeInput}
\end{CodeChunk}

\begin{figure}
    \centering
    \includegraphics[width=0.95\textwidth]{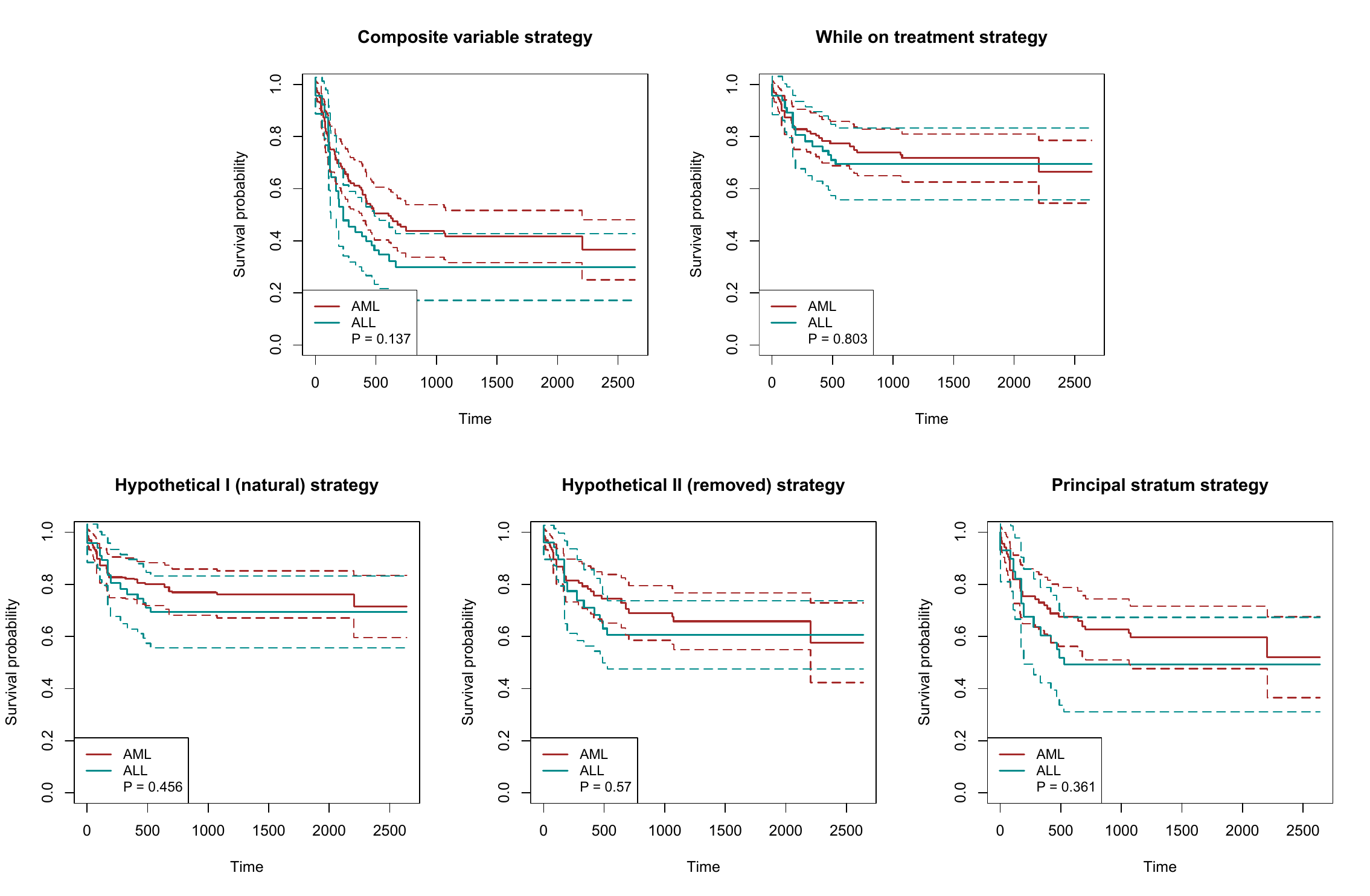}
    \caption{Analysis 2 for BMT data: survival functions. Competing risks data, semiparametrically efficient estimation.} \label{fig:bmt2}
\end{figure}

\begin{figure}
    \centering
    \includegraphics[width=0.95\textwidth]{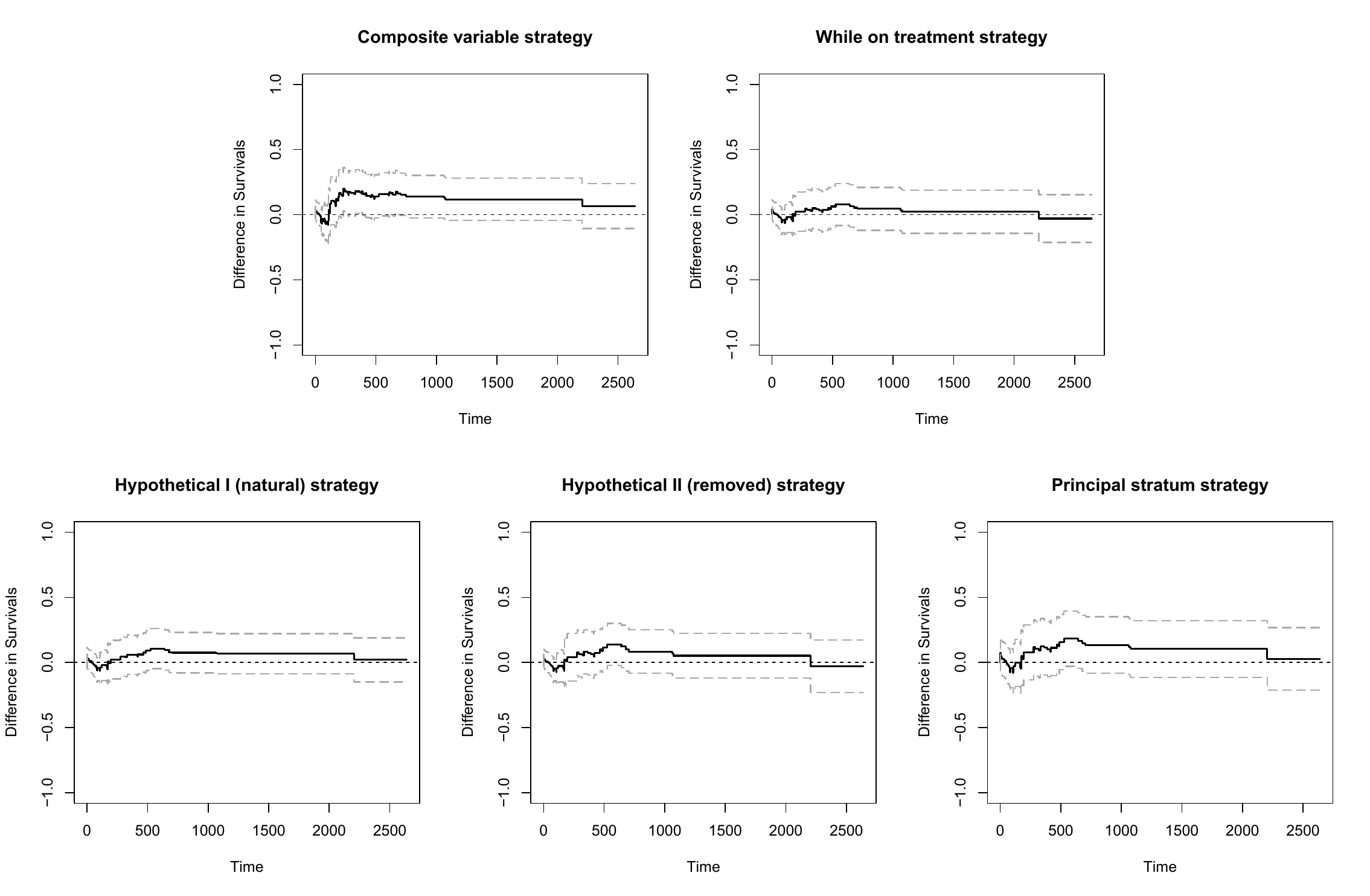}
    \caption{Analysis 2 for BMT data: treatment effects. Competing risks data, semiparametrically efficient estimation.} \label{fig:bmt2_ate}
\end{figure}

Figure \ref{fig:bmt2} shows the estimated survival functions under AML and ALL using semiparametrically efficient estimation. Figure \ref{fig:bmt2_ate} shows the estimated treatment effects. The treatment effect is the difference in survival functions between AML and ALL.


\subsection{Analysis in the semi-competing risks setting}

The semi-competing risks data include the death time \code{t1} with indicator \code{d1} (1 for death, 0 for censoring) and the relapse time \code{t2} with indicator \code{d2} (1 for relapse, 0 for censoring). 
We use nonparametric estimation with inverse treatment probability weighting (based on a non-standardized propensity score). From a statistical perspective, the standard error may be inaccurate because the uncertainty of the estimated propensity score is not accounted for in the inference. Therefore, we use bootstrapping to construct confidence intervals from 200 resamples.

{\begin{CodeChunk}
\begin{CodeInput}
R> slist = c("treatment", "composite", "whileon", "natural", "removed", 
R>           "principal")
R> for (st in slist){
R>   fit = scr.tteICE(A, bmt$t1, bmt$d1, bmt$t2, bmt$d2, st, X, 
R>                    method = "ipw", seed = 10, nboot = 200)
R>   plot(fit, type = "inc", plot.configs = list(legend = c('AML', 'ALL')))
R>   plot(fit, type = "ate")
R> }
\end{CodeInput}
\end{CodeChunk}}

\begin{figure}
    \centering
    \includegraphics[width=0.95\textwidth]{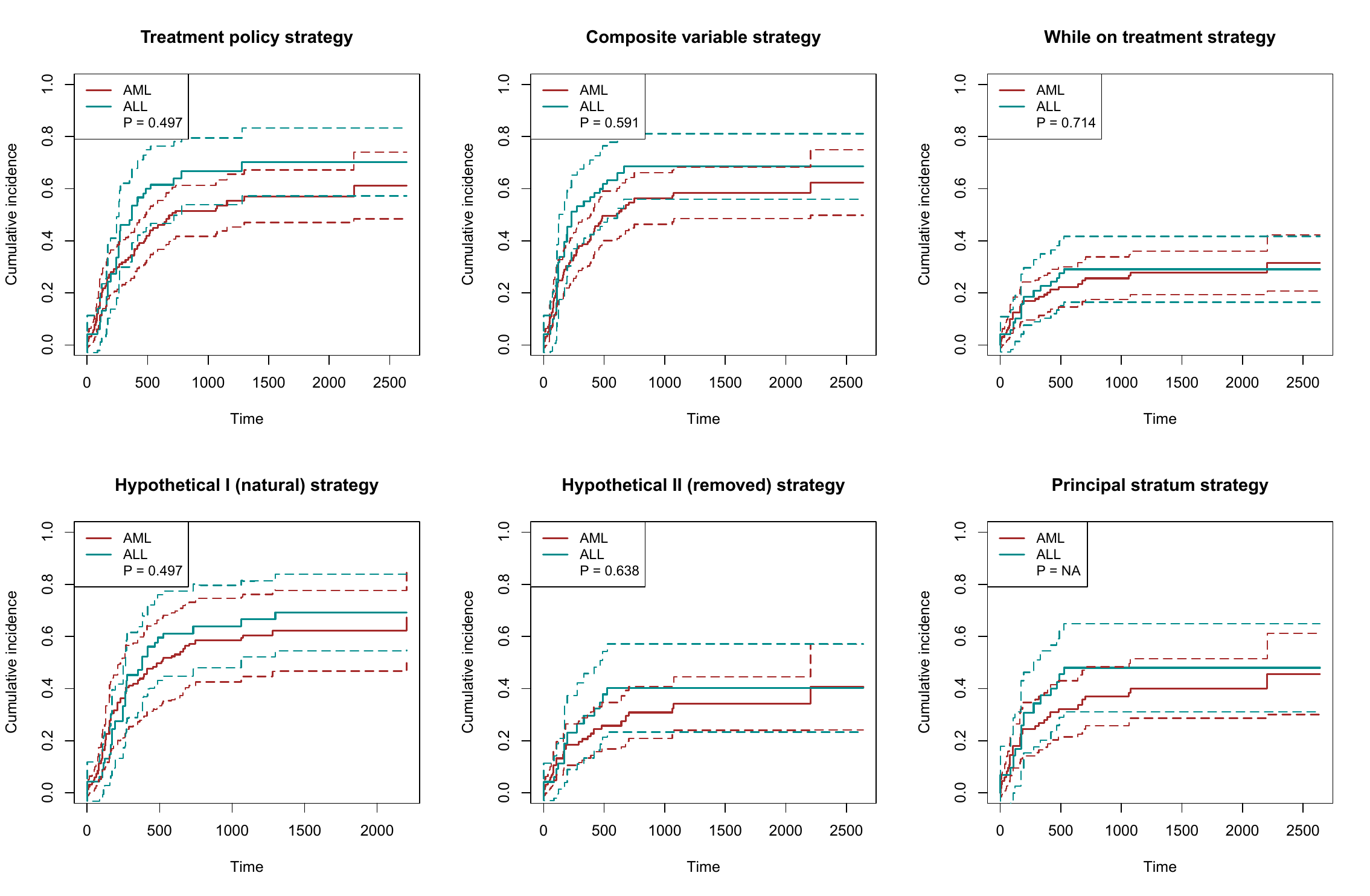}
    \caption{Analysis 3 for BMT data: cumulative incidence functions. Semi-competing risks data, nonparametric estimation with inverse treatment probability weighting.} \label{fig:bmt3}
\end{figure}

\begin{figure}
    \centering
    \includegraphics[width=0.95\textwidth]{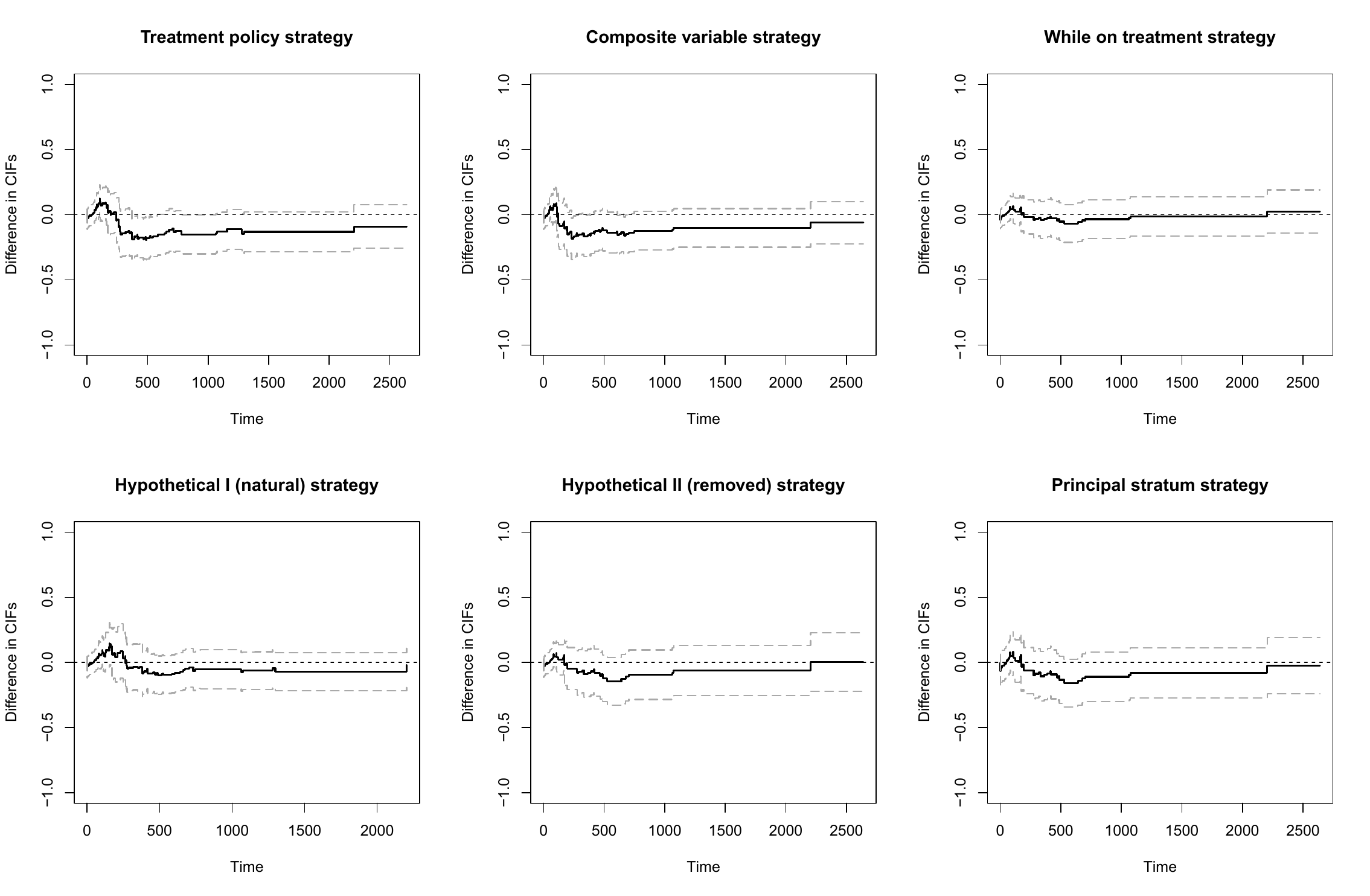}
    \caption{Analysis 3 for BMT data: treatment effects. Semi-competing risks data, nonparametric estimation with inverse treatment probability weighting.} \label{fig:bmt3_ate}
\end{figure}

Figure \ref{fig:bmt3} shows the estimated cumulative incidence functions under AML and ALL using nonparametric estimation with inverse treatment probability weighting. Figure \ref{fig:bmt3_ate} shows the estimated treatment effects. The treatment effect is the difference in cumulative incidence functions between AML and ALL. Although $p$-values are available for all strategies for semi-competing risks data, we do not display them because they may not be accurate due to inverse treatment probability weighting.


Finally, we perform the semiparametrically efficient estimation. We select a subsample of patients older than 15 years.

\begin{CodeChunk}
\begin{CodeInput}
R> subs = (X[, 1] >= 15)
R> slist = c("treatment", "composite", "whileon", "natural", "removed", 
R>           "principal")
R> for (st in slist){
R>   fit = scr.tteICE(A, bmt$t1, bmt$d1, bmt$t2, bmt$d2, st, method = "eff", 
R>                    X, subset = subs)
R>   plot(fit, type = "inc", plot.configs = list(legend = c('AML', 'ALL')))
R>   plot(fit, type = "ate")
R> }
\end{CodeInput}
\end{CodeChunk}

\begin{figure}
    \centering
    \includegraphics[width=0.95\textwidth]{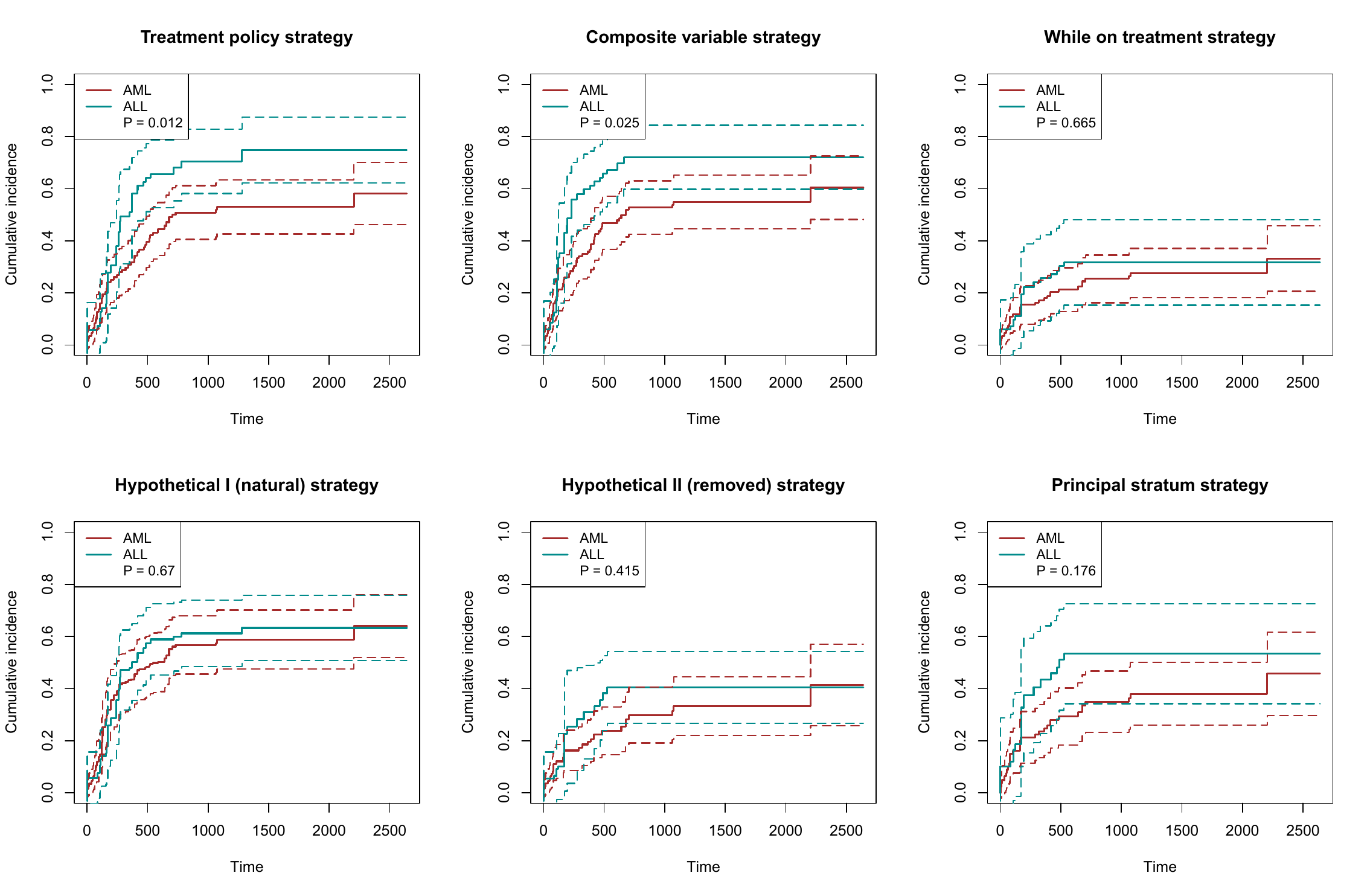}
    \caption{Analysis 4 for BMT data: cumulative incidence functions. Semi-competing risks data, semiparametrically efficient estimation in a subset.} \label{fig:bmt4}
\end{figure}

\begin{figure}
    \centering
    \includegraphics[width=0.95\textwidth]{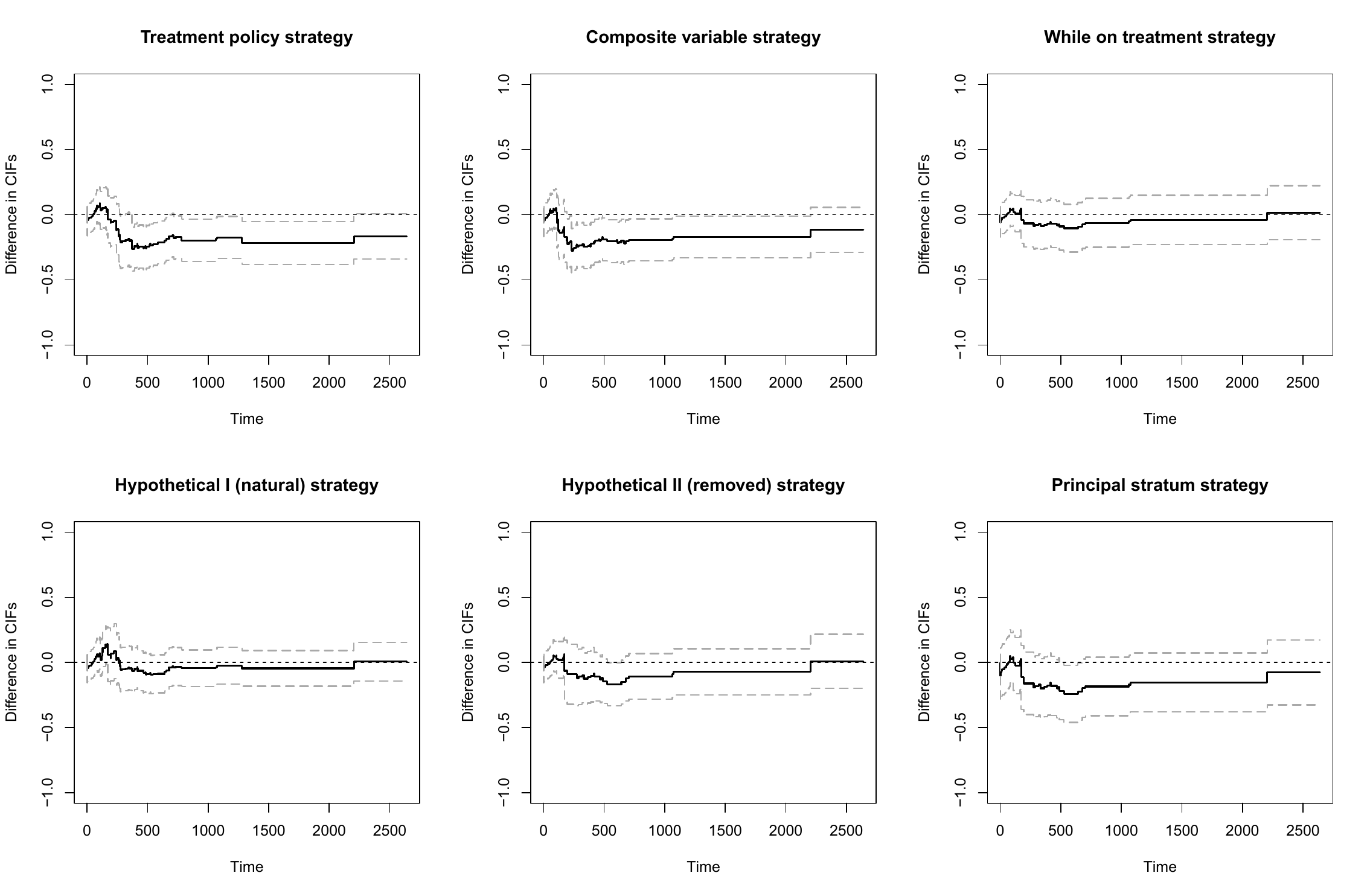}
    \caption{Analysis 4 for BMT data: treatment effects. Semi-competing risks data, semiparametrically efficient estimation in a subset.} \label{fig:bmt4_ate}
\end{figure}

Figure \ref{fig:bmt4} shows the estimated cumulative incidence functions under AML and ALL using semiparametrically efficient estimation in the subset. Figure \ref{fig:bmt4_ate} shows the estimated treatment effects. The treatment effect is the difference in cumulative incidence functions between AML and ALL. The treatment effect is significant in the treatment policy strategy, indicating that the mortality risk is lower in the AML group. Note that this trial was an observational study. All three covariates have different distributions between groups, as verified by $t$-tests and $\chi^2$-tests. Therefore, the analysis results could be more reliable with semiparametrically efficient estimation. The estimated functions are consistent if either the hazard models or the propensity score model is correctly specified.


According to Figures \ref{fig:bmt3} and \ref{fig:bmt4}, the while on treatment strategy yields the lowest cumulative incidence functions. This is because all death events after relapse are excluded, resulting in the fewest individuals reaching endpoints. The hypothetical II strategy (controlled effect) yields lower cumulative incidence functions than the hypothetical I strategy (natural effect). This is because relapse is associated with an increased risk of death, and thus removing relapse can reduce the risk of death. The principal stratum strategy shows wider confidence intervals because the target population is smaller and there is greater variation when estimating the proportion in the target principal stratum.

Finally, we predict the risk at specific time points. Suppose that we use \fct{tteICE} to fit a model under the treatment policy strategy using semiparametrically efficient estimation. The estimates and parameters are presented by the \fct{print} function. The \fct{summary} function additionally outputs the predicted cumulative incidence functions and the treatment effects at quartiles of follow-up time. The \fct{predict} function gives the cumulative incidence functions and treatment effects. The treatment effects are negative, indicating that the AML group has a lower mortality rate.

\begin{CodeChunk}
\begin{CodeInput}
R> bmt$A = A
R> fit = tteICE(Surv(t1, d1) ~ A | z1 + z3 + z5, add.scr = ~ Surv(t2, d2), 
R>              data = bmt, strategy = "treatment", method = "eff")
R>
R> print(fit, digits = 3) 
# Input:
# tteICE(formula = Surv(t1, d1) ~ A | z1 + z3 + z5, add.scr = ~Surv(t2, 
#     d2), data = bmt, strategy = "treatment", method = "eff")
# -----------------------------------------------------------------------
# Data type: semicompeting risks 
# Strategy: treatment policy strategy 
# Estimation method: semiparametrically efficient estimation 
# Observations: 137 (including 99 treated and 38 control)
# Maximum follow-up time: 2640 
# P-value of the average treatment effect: 0.064 
R>
R> summary(fit)
# Input:
# tteICE(formula = Surv(t1, d1) ~ A | z1 + z3 + z5, add.scr = ~Surv(t2, 
#     d2), data = bmt, strategy = "treatment", method = "eff")
# -----------------------------------------------------------------------
# Data type: semicompeting risks 
# Strategy: treatment policy strategy 
# Estimation method: semiparametrically efficient estimation 
# Observations: 137 (including 99 treated and 38 control)
# Maximum follow-up time: 2640 
# P-value of the average treatment effect: 0.064 
# -----------------------------------------------------------------------
# -----------------------------------------------------------------------
# The estimated cumulative incidences and treatment effects at quartiles:
#          660   1320   1980   2640
# CIF1   0.485  0.570  0.570  0.616
# se1    0.052  0.052  0.052  0.058
# CIF0   0.635  0.732  0.732  0.732
# se0    0.068  0.068  0.068  0.068
# ATE   -0.150 -0.162 -0.162 -0.116
# se     0.085  0.086  0.086  0.090
# p.val  0.078  0.059  0.059  0.194
R> 
R> round(predict(fit, 365*(1:5)), digits=3)
#          365    730   1095   1460   1825
# CIF1   0.344  0.507  0.537  0.570  0.570
# se1    0.049  0.052  0.052  0.052  0.052
# CIF0   0.499  0.658  0.683  0.732  0.732
# se0    0.080  0.066  0.065  0.068  0.068
# ATE   -0.155 -0.150 -0.146 -0.162 -0.162
# se     0.094  0.084  0.083  0.086  0.086
# p.val  0.100  0.074  0.079  0.059  0.059
\end{CodeInput}
\end{CodeChunk}

In this example, the estimated baseline hazards were extracted as follows. The first column represents the time points, and the subsequent columns show the estimated baseline cumulative hazards from the Cox models.

\begin{CodeChunk}
\begin{CodeInput}
R> head(bshaz(fit))
#   time     cumhaz1    cumhaz0
# 1    0 0.000000000 0.00000000
# 2    1 0.000000000 0.02887603
# 3    2 0.009560267 0.02887603
# 4   10 0.019188625 0.02887603
# 5   16 0.028914054 0.02887603
# 6   35 0.038767098 0.02887603
\end{CodeInput}
\end{CodeChunk}

The estimated coefficients for each group are shown below. For example, the estimated log hazard ratio of the covariate \code{z3} is $-0.2986$ in the Cox model for the survival outcome in the treated group and $0.0385$ in the Cox model for the survival outcome in the control group. If users want to construct hypothetical survival functions for a given covariate value, the results from \fct{bshaz} and \fct{coef} will be used.

\begin{CodeChunk}
\begin{CodeInput}
R> coef(fit)
#    Primary, A=1         SE Primary, A=0         SE
# z1  0.006875631 0.01398301   0.07396413 0.03335849
# z3 -0.298624078 0.26588480   0.03852526 0.47849066
# z5  0.241233914 0.27265205  -0.39030496 0.45834578
\end{CodeInput}
\end{CodeChunk}

The proportional hazards assumption in the working Cox model is tested based on the Schoenfeld residuals. Since the treatment policy is used, we fitted two Cox models for the primary outcome event in the treated and control groups, and \code{zph(fit)} outputs the $p$-values of the tests in each group, for each covariate and globally. The proportional hazards assumption is not rejected in the global sense. However, the effect of the covariate \code{z5} on the hazard in the control group may not be proportional, so the coefficient $-0.3903$ needs to be interpreted as a weighted log hazard ratio.

\begin{CodeChunk}
\begin{CodeInput}
R> zph(fit)
#        Primary, A=1 Primary, A=0
# z1        0.3850827   0.16125665
# z3        0.7008233   0.18759931
# z5        0.6862000   0.03943241
# GLOBAL    0.7648817   0.12471514
\end{CodeInput}
\end{CodeChunk}

\section{Summary and discussion} \label{sec:summary}

A formalized workflow for analyzing time-to-event outcomes is crucial, particularly when addressing intercurrent events. Existing practices, such as using \fct{coxph} in the \pkg{survival} package or \fct{crr} in the \pkg{cmprsk} package, often constitute \textit{ad hoc} analyses that yield only an estimated treatment hazard ratio. While informative, these methods do not formally align with the estimand framework defined by ICH E9 (R1). For comparison, our \pkg{tteICE} incorporates these naive analyses in the \fct{surv.tteICE} function, allowing researchers to directly contrast traditional and estimand-based approaches.

The \pkg{tteICE} package provides a unified framework for estimating cumulative incidence functions and treatment effects in accordance with ICH E9 (R1). Although theoretical formulations have been proposed under the competing risks data structure \citep{deng2025inference}, implementations can be difficult for practitioners. This package implements theoretical derivations in practice and supports common data structures, including both competing and semi-competing risks, across diverse study designs. This package incorporates both nonparametric and semiparametrically efficient estimation methods, offering users the flexibility to specify individual weights or select subsets for analysis. The implementation is also computationally efficient and compatible, as it is built on established, optimized \proglang{R} packages. To enhance accessibility, we provide an interactive Shiny application that guides users through data upload, descriptive analysis, estimation and inference, and results visualization.

The \pkg{tteICE} package facilitates a principled analytical process for practitioners analyzing randomized controlled trial and observational study data. In the design phase, users first determine their data structure (competing or semi-competing risks) and then select the appropriate ICH E9 (R1) strategy for handling ICEs. In the analysis phase, users determine the estimation method. Nonparametric estimation is often advantageous in randomized controlled trials due to its robustness. On the other hand, semiparametrically efficient estimation, which leverages covariate information to achieve asymptotic efficiency and robustness to internal model misspecification, is particularly beneficial in observational studies. However, this efficiency comes at a computational cost, as it involves fitting covariate-specific hazards, especially intensive for the hypothetical strategy (Scenario I, natural), which requires modeling additional hazard functions. Finally, this package can easily produce plots of cumulative incidence functions and time-varying treatment effects with corresponding confidence intervals.

Some potential extensions of the \pkg{tteICE} package are under consideration but remain challenging to implement at present. First, extending the package to accommodate additional data types and estimands is of interest. A common scenario in clinical trials is where the occurrence of an ICE is known, but its exact timing is unobservable. While the five strategies remain well-defined in this case, estimation may require alternative approaches, such as imputation or mixture models, for which methodological developments are still ongoing. As these statistical methods mature, they will be integrated into the package.
Second, we aim to enhance the inference capabilities. Currently, the package provides pointwise inference and a universal $p$-value. Future updates could support inference on other clinically relevant estimands, such as the average hazard ratio, as these methodologies gain broader recognition and acceptance within the statistical community.
Third, more flexible working models than the Cox model can be incorporated into semiparametrically efficient estimation, which requires additional effort to diagnose and compare fitted models. While the interpretability of the estimated conditional hazards from nonparametric or machine-learning methods is less straightforward, the proportional hazards assumption and Markovness can be relaxed, thereby avoiding bias due to model misspecification.

\section*{Acknowledgments}

The authors thank the journal editor and anonymous reviewers for their comments, which helped us improve the computational efficiency and readability of our package.

\section*{Funding information}
Y.Z. is supported by Grant-in-Aid for Early-Career Scientists (Grant No.~26K21181) and Grant-in-Aid for Scientific Research (Grant No.~25K03086) from the Ministry of Education, Science, Sports and Technology in Japan. Y.W. is supported by the Young Scientists Fund of the National Natural Science Foundation of China (Grant No.~12401358). S.H. is supported by the National Natural Science Foundation of China (Grant No.~82304269), the Chinese Academy of Medical Sciences Innovation Fund for Medical Sciences (Grant No.~2023-I2M-3-008), and the Key Research and Development Program of Jiangxi Province (Grant No.~20244AFI92004).

\bibliography{jss6272}
\clearpage

\newpage

\begin{appendix}

\setcounter{table}{0}
\renewcommand{\thetable}{S\arabic{table}}
\setcounter{figure}{0}
\renewcommand{\thefigure}{S\arabic{figure}}

\section*{APPENDIX}

\section{More technical details} \label{sec:technical}

In the following, we list the estimators and their estimated asymptotic variances (avars) for all strategies. Given an estimator $\widehat\theta$ for $\theta$, the standard error is calculated as $\mathrm{se}(\widehat\theta) = \{\widehat\avar(\widehat\theta)/n\}^{1/2}$, where $n$ is the sample size.

\subsection{Nonparametric estimation}

The asymptotic variances of the CIF and treatment effect estimators are derived using the functional delta method, which involves Hadamard derivatives with respect to cumulative hazard functions. Theoretical derivations of the standard errors can be found in a previous work \citep{deng2025inference}.

\paragraph{Treatment policy strategy}

\begin{align*}
\widehat\mu_w^{\text{tp}}(t) & = 1- \exp\{-\widehat\Lambda(t;w)\}, \\
\widehat\tau^{\text{tp}}(t) &= \widehat\mu_1^{\text{tp}}(t) - \widehat\mu_0^{\text{tp}}(t), \\
\widehat\avar\{\widehat\mu_w^{\text{tp}}(t)\} &= \exp\{-2\widehat\Lambda(t;w)\} \int_0^t \frac{d\widehat\Lambda(s;w)}{\widehat{{P}}(\tilde{T} \ge s, W=w)}, \\
\widehat\avar\{\widehat\tau^{\text{tp}}(t)\} &= \widehat\avar\{\widehat\mu_1^{\text{tp}}(t)\} + \widehat\avar\{\widehat\mu_0^{\text{tp}}(t)\},
\end{align*}

\paragraph{Composite variable strategy}

\begin{align*}
\widehat\mu_w^{\text{cv}}(t) & = 1- \exp\{-\widehat\Lambda_{12}(t;w)\}, \\
\widehat\tau^{\text{cv}}(t) &= \widehat\mu_1^{\text{cv}}(t) - \widehat\mu_0^{\text{cv}}(t), \\
\widehat\avar\{\widehat\mu_w^{\text{cv}}(t)\} &= \exp\{-2\widehat\Lambda_{12}(t;w)\} \int_0^t \frac{d\widehat\Lambda_{12}(s;w)}{\widehat{{P}}(\tilde{T} \ge s, W=w)}, \\
\widehat\avar\{\widehat\tau^{\text{cv}}(t)\} &= \widehat\avar\{\widehat\mu_1^{\text{cv}}(t)\} + \widehat\avar\{\widehat\mu_0^{\text{cv}}(t)\},
\end{align*}

\paragraph{While on treatment strategy}

\begin{align*}
\widehat\mu_w^{\text{wo}}(t) &= \int_0^t \exp\{-\widehat\Lambda_{12}(s;1)\}d\widehat\Lambda_1(s;1), \\
\tau^{\text{wo}}(t) &= \widehat\mu_1^{\text{wo}}(t) - \widehat\mu_0^{\text{wo}}(t), \\
\widehat\avar\{\widehat\mu_w^{\text{wo}}(t)\} &=
\int_0^t \bigg[ \{e^{-\widehat\Lambda_{12}(s;w)}-\widehat\mu_w^{\text{wo}}(t)+\widehat\mu_w^{\text{wo}}(s)\}^2 \frac{d\widehat\Lambda_1(s;w)}{\widehat{{P}}(\tilde{T}\wedge\tilde{R} \ge s, W=w)} \\
&\qquad\quad + \{\widehat\mu_w^{\text{wo}}(t)-\widehat\mu_w^{\text{wo}}(s)\}^2 \frac{d\widehat\Lambda_2(s;w)}{{P}(\tilde{T}\wedge\tilde{R} \ge s, W=w)} \bigg], \\
\widehat\avar\{\widehat\tau^{\text{wo}}(t)\} &= \widehat\avar\{\widehat\mu_1^{\text{wo}}(t)\} + \widehat\avar\{\widehat\mu_0^{\text{wo}}(t)\}.
\end{align*}

\paragraph{Hypothetical strategy I (competing risks)}

\begin{align*}
\widehat\mu_w^{\text{hp,I}}(t)
&= \int_0^t \exp\{-\widehat\Lambda_1(s;w) - \widehat\Lambda_2(s;0)\}d\widehat\Lambda_1(s;w), \\
\widehat\tau^{\text{hp,I}}(t) &= \widehat\mu_1^{\text{hp,I}}(t) - \widehat\mu_0^{\text{hp,I}}(t), \\
\widehat\avar\{\widehat\mu_w^{\text{hp,I}}(t)\}
&= \int_0^t \{e^{-\widehat\Lambda_1(s;w)-\widehat\Lambda_2(s;0)}-\widehat\mu_w^{\text{hp,I}}(t)+\widehat\mu_w^{\text{hp,I}}(s)\}^2 \frac{d\widehat\Lambda_1(s;w)}{\widehat{{P}}(\tilde{T}\wedge\tilde{R} \ge s, W=1)} \\
&\qquad + \int_0^t \{\widehat\mu_w^{\text{hp,I}}(t)-\mu_w^{\text{hp,I}}(s)\}^2 \frac{d\widehat\Lambda_2(s;0)}{\widehat{{P}}(\tilde{T}\wedge\tilde{R} \ge s, W=0)}, \\
\widehat\avar\{\widehat\tau^{\text{hp,I}}(t)\}
&= \int_0^t \{e^{-\widehat\Lambda_1(s;1)-\widehat\Lambda_2(s;0)}-\widehat\mu_1^{\text{hp,I}}(t)+\widehat\mu_1^{\text{hp,I}}(s)\}^2 \frac{d\widehat\Lambda_1(s;1)}{\widehat{{P}}(\tilde{T}\wedge\tilde{R} \ge s, W=1)} \\
&\qquad + \int_0^t \{e^{-\widehat\Lambda_1(s;0)-\widehat\Lambda_2(s;0)}-\widehat\mu_0^{\text{hp,I}}(t)+\widehat\mu_0^{\text{hp,I}}(s)\}^2 \frac{d\widehat\Lambda_1(s;0)}{\widehat{{P}}(\tilde{T}\wedge\tilde{R} \ge s, W=0)} \\
&\qquad + \int_0^t \{\widehat\mu_1^{\text{hp,I}}(t)-\widehat\mu_0^{\text{hp,I}}(t)-\widehat\mu_1^{\text{hp,I}}(s)+\widehat\mu_0^{\text{hp,I}}(s)\}^2 \frac{d\widehat\Lambda_2(s;0)}{\widehat{{P}}(\tilde{T}\wedge\tilde{R} \ge s, W=0)}.
\end{align*}

\paragraph{Hypothetical strategy I (semi-competing risks)}

\begin{align*}
\widehat\mu_w^{\text{hp,I}}(t) &= 1 - \exp\{-\widehat\Lambda_1(t;w,0)-\widehat\Lambda_2(t;0)\} \\
&\quad - \int_0^t\exp\{-\widehat\Lambda_1(s;w,0)-\widehat\Lambda_2(s;0)-\widehat\Lambda_1(t;w,1)+\widehat\Lambda_1(s;w,1)\}d\widehat\Lambda_2(s;0), \\
\widehat{\avar}\{\widehat\tau^{\text{hp,I}}(t)\} &= \widehat\mu_1^{\text{hp,I}}(t) - \widehat\mu_0^{\text{hp,I}}(t), \\
\widehat\mu_w^{\text{hp,I}}(t) &= \int_0^t \{1-\widehat\mu_w^{\text{hp,I}}(t)-C(s,t,w)\}^2 \frac{d\Lambda_1(s;w,0)}{{P}(\tilde{T}\wedge\tilde{R} \geq s, W=w)} \\
&\quad+ \int_0^t \{1-\widehat\mu_w^{\text{hp,I}}(t)-D(s,t,w)\}^2 \frac{d\widehat\Lambda_2(s;0)}{\widehat{{P}}(\tilde{T}\wedge\tilde{R}\geq s, W=0)} \\
&\quad+ \int_0^t C(s,t,w)^2 \frac{d\widehat\Lambda_1(s;w,1)}{\widehat{{P}}(\tilde{T}\geq s, \tilde{R}\leq s, W=w)}, \\
\widehat{\avar}\{\widehat\tau^{\text{hp,I}}(t)\} &= \int_0^t \{1-\widehat\mu_1^{\text{hp,I}}(t)-C(s,w,1)\}^2 \frac{d\Lambda_1(s;1,0)}{{P}(\tilde{T}\wedge\tilde{R} \geq s, W=1)} \\
&\quad + \int_0^t \{1-\widehat\mu_0^{\text{hp,I}}(t)-C(s,w,0)\}^2 \frac{d\Lambda_1(s;0,0)}{{P}(\tilde{T}\wedge\tilde{R} \geq s, W=0)} \\
&\quad + \int_0^t \{\widehat\mu_0^{\text{hp,I}}(t)-\widehat\mu_1^{\text{hp,I}}(t)+D(s,t,0)-D(s,t,1)\}^2 \frac{d\widehat\Lambda_2(s;0)}{\widehat{{P}}(\tilde{T}\wedge\tilde{R}\geq s, W=0)} \\
&\quad + \int_0^t \left[\frac{C(s,t,1)^2 d\widehat\Lambda_1(s;1,1)}{\widehat{{P}}(\tilde{T}\geq s, \tilde{R}\leq s, W=1)} + \frac{C(s,t,0)^2 d\widehat\Lambda_1(s;0,1)}{\widehat{{P}}(\tilde{T}\geq s, \tilde{R}\leq s, W=0)}\right],
\end{align*} 
where
\begin{align*}
C(s,t,w) &= \int_0^s e^{-\widehat\Lambda_1(u;w,0)-\widehat\Lambda_2(u;0)-\Lambda_1(t;w,1)}\{e^{\widehat\Lambda_1(s;w,1)}-e^{\widehat\Lambda_1(u;w,1)}\} d\widehat\Lambda_2(s;0), \\
D(s,t,w) &= \{1-\widehat\mu_w^{\text{hp,I}}(s)\} e^{\widehat\Lambda_1(s;w,1)-\Lambda_1(t;w,1)}.
\end{align*}

\paragraph{Hypothetical strategy II}

\begin{align*}
\widehat\mu_w^{\text{hp,II}}(t)
&= 1 - \exp\{-\widehat\Lambda_1(s;w)\}, \\
\widehat\tau^{\text{hp,I}}(t) &= \widehat\mu_1^{\text{hp,II}}(t) - \widehat\mu_0^{\text{hp,II}}(t), \\
\widehat\avar\{\widehat\mu_w^{\text{hp,II}}(t)\}
&= \exp\{-2\widehat\Lambda_1(t;w)\} \int_0^t \frac{d\widehat\Lambda_1(s;w)}{\widehat{{P}}(\tilde{T}\wedge\tilde{R} \ge s, W=w)}, \\
\widehat\avar\{\widehat\tau^{\text{hp,II}}(t)\} &= \widehat\avar\{\widehat\mu_1^{\text{hp,II}}(t)\} + \widehat\avar\{\widehat\mu_0^{\text{hp,II}}(t)\}.
\end{align*}

\paragraph{Principal stratum strategy}

\begin{align*}
\widehat\mu_w^{\text{ps}}(t) &= \frac{\int_0^t \exp\{-\widehat\Lambda_{12}(s;w)\} d\widehat\Lambda_1(s;w)}{1 - \int_0^{t^*} \exp\{-\widehat\Lambda_{12}(s;w)\} d\widehat\Lambda_2(s;w)}, \\
\widehat\tau^{\text{ps}}(t) &= \widehat\mu_1^{\text{ps}}(t) - \widehat\mu_0^{\text{ps}}(t), \\
\widehat\avar\{\widehat\mu_w^{\text{ps}}(t)\}
&= \frac{\int_0^{t^*} \{A_1(s,t,w)-\widehat\mu_w^{\text{ps}}(t)A_2(s,t^*,w)\}^2 \widehat{{P}}(\tilde{T}\wedge\tilde{R} \ge s, W=w)^{-1}d\widehat\Lambda_1(s;w)}{\{1 - \int_0^{t^*} e^{-\widehat\Lambda_{12}(s;w)} d\widehat\Lambda_2(s;w)\}^2} \\
&\quad + \frac{\int_0^{t^*} \{B_1(s,t,w)-\widehat\mu_w^{\text{ps}}(t)B_2(s,t^*,w)\}^2 \widehat{{P}}(\tilde{T}\wedge\tilde{R} \ge s, W=w)^{-1}d\widehat\Lambda_2(s;w)}{\{1 - \int_0^{t^*} e^{-\widehat\Lambda_{12}(s;w)} d\widehat\Lambda_2(s;w)\}^2}, \\
\widehat\avar\{\widehat\tau^{\text{ps}}(t)\} &= \widehat\avar\{\widehat\mu_1^{\text{ps}}(t)\} + \widehat\avar\{\widehat\mu_0^{\text{ps}}(t)\},
\end{align*}
where
\begin{align*}
A_1(s,t,w) &= [\exp\{-\widehat\Lambda_{12}(s;w)\}+\widehat\mu_w^{\text{wo}}(s)-\widehat\mu_w^{\text{wo}}(t)] I(s \le t), \\
A_2(s,t,w) &= \exp\{-\widehat\Lambda_{12}(s;w)\}-\exp\{-\widehat\Lambda_{12}(t;w)\}+\widehat\mu_w^{\text{wo}}(s)-\widehat\mu_w^{\text{wo}}(t), \\
B_1(s,t,w) &= \{\widehat\mu_w^{\text{wo}}(t)-\widehat\mu_w^{\text{wo}}(s)\} I(s \le t), \\
B_2(s,t,w) &= \exp\{-\widehat\Lambda_{12}(t;w)\}+\widehat\mu_w^{\text{wo}}(t)-\widehat\mu_w^{\text{wo}}(s).
\end{align*}

\subsection{Semiparametrically efficient estimation}

Suppose the working models are correctly specified. EIFs are additive, so the EIFs of the treatment effects are contrasts of EIFs of the CIFs under treatment and control. According to empirical process theory, the asymptotic variances of the CIF estimators are the variances of their associated EIFs \citep[Supplementary Material]{deng2025inference}. Let $\widehat\varphi_w^{k}(t|X)$ be the estimated non-centered EIF of the cumulative incidence function associated with treatment condition $w$ under strategy $k$. Let $P_n$ be the sample average operator. Then 
\begin{align*}
\widehat\mu_w^{k}(t)& = {P}_n \{\widehat\varphi_w^{k}(t|X)\}, \\
\widehat\tau^{k}(t) &= \widehat\mu_1^{k}(t) - \widehat\mu_0^{k}(t), \\
\widehat{\avar}\{\widehat\mu_w^{k}(t)\} &= {P}_n \{\widehat\varphi_w^{k}(t|X)^2\}, \\
\widehat{\avar}\{\widehat\tau^{k}(t)\} &= {P}_n [\{\widehat\varphi_1^{k}(t|X)-\widehat\varphi_1^{k}(t|X)\}^2].
\end{align*}

\section{Simulation-based validation of the package}\label{sec:simulation}

Consider a completely randomized trial. Suppose that the hazard specific to the potential primary outcome event and intercurrent event are $\lambda_1(t;w) = a_w t$ and $\lambda_2(t;w) = c_w$, $w = 1, 0$, respectively. Therefore, the marginal distribution of the potential primary outcome event is Weibull, with $T(w) \sim Weibull(2, \sqrt{2/a_w})$. Suppose the intercurrent event neither prevents nor modifies the hazard of the primary outcome event. We collect data in the competing risks structure. Table \ref{tab:simu1} lists the explicit forms of $\mu_w^k(t)$. 

\begin{table}
\centering
\caption{The explicit form of the cumulative incidence functions in the simulation with competing risks data} \label{tab:simu1}
\begin{tabular}{ll}
\toprule
Strategy & $\mu_w^k (t)$, $0 \le t \le t^*$ \\ \midrule
tp & $1 - e^{-a_w t^2/2}$ \\
cv & $1 - e^{-a_w t^2/2 - c_w t}$ \\
wo & $1-e^{-a_w t^2/2 - c_w t} - e^{c_w^2/2a_w} (2\pi c_w^2/a_w)^{1/2}\{\Phi(a_w^{1/2}(t+c_w/a_w)) - \Phi(c_w/a_w^{1/2})\}$ \\
hp,I & $1-e^{-a_w t^2/2 - c_0 t} - e^{c_0^{2}/2a_w} (2\pi c_0^{2}/a_w)^{1/2}\{\Phi(a_w^{1/2}(t+c_0/a_w)) - \Phi(c_0/a_w^{1/2})\}$ \\
hp,II & $1 - e^{-a_w t^2/2}$ \\
ps & $\frac{1-e^{-a_w t^2/2 - c_w t} - e^{c_w^2/2a_w} (2\pi c_w^2/a_w)^{1/2}\{\Phi(a_w^{1/2}(t+c_w/a_w)) - \Phi(c_w/a_w^{1/2})\}}{1 - e^{c_w^2/2a_w} {(2\pi c_w^2/a_w)}^{1/2}\{\Phi(a_w^{1/2}(t^*+c_w/a_w)) - \Phi(c_w/a_w^{1/2})\}}$ \\
\bottomrule
\end{tabular} \\
Notes: $\Phi(\cdot)$ denotes the cumulative distribution function of the standard normal distribution.
\end{table}

We simulate data $B=1000$ times and evaluate the empirical performance of nonparametric estimation. We set the sample size \code{n = 500}. In each replicate, the treatment assignment is generated from \code{rbinom(n, 1, 0.5)}. We set the hazards as follows.
\begin{Code}
  a1 = 0.05; a0 = 0.03; c1 = 0.04; c0 = 0.05
\end{Code}
We use \code{rweibull(n, 2, sqrt(2 / a))} to generate the time to the primary outcome event and use \code{rexp(N, c)} to generate the time to the intercurrent event, where \code{a} represents $\{a_1, a_0\}$ and \code{c} represents $\{c_1, c_0\}$. Censoring is generated from the minimum of a uniform distribution \code{runif(n, 4, 8)} and $7$. 

\begin{CodeChunk}
\begin{CodeInput}
R> generatedata <- function(n){
R>   A = rbinom(n, 1, 0.5)
R>   T1 = rweibull(n, 2, sqrt(2 / a1))
R>   T0 = rweibull(n, 2, sqrt(2 / a0))
R>   R1 = rexp(n, c1)
R>   R0 = rexp(n, c0)
R>   C = runif(n, 4, 8)
R>   C[C > 7] = 7
R>   T = T1 * A + T0 * (1 - A)
R>   R = R1 * A + R0 * (1 - A)
R>   R[R >= T] = 99
R>   dT = as.numeric(T <= C)
R>   dR = as.numeric(R <= C)
R>   T = T * dT + C * (1 - dT)
R>   R = R * dR + C * (1 - dR)
R>   Time = (T + R - abs(T - R)) / 2
R>   cstatus = dT + 2 * dR
R>   cstatus[cstatus > 2] = 2
R>   return(list(A = A, T = T, R = R, dT = dT, dR = dR, 
R>          Time = Time, cstatus = cstatus))
R> }
\end{CodeInput}
\end{CodeChunk}

For the treatment policy strategy, \code{A}, \code{T}, and \code{dT} from the output of \fct{generatedata} will be used. For other strategies, \code{A}, \code{Time}, and \code{cstatus} will be used. To facilitate replication of the results, we separately evaluate the asymptotic standard error (confidence interval) and the bootstrap standard error (confidence interval) from 200 resamples. Finishing the simulation for the asymptotic standard error (confidence interval) takes less than five minutes on a Linux server with Intel(R) Xeon(R) Gold 6140, 2.30GHz CPU, whereas bootstrapping takes about ten hours. Table \ref{tab:simu1_bias} shows the empirical bias of estimated treatment effects, standard deviation of estimated treatment effects among $B=1000$ replicates, mean standard error of estimated treatment effects, and coverage percentage of 95\% confidence intervals at selected time points. The bias is negligible. The standard error is close to the standard deviation. The confidence intervals have coverage rates close to the nominal level. The asymptotic standard errors (confidence intervals) are similar to bootstrap standard errors (confidence intervals), suggesting that using the asymptotic formula is statistically valid and computationally efficient for inference.

\begin{table}
\centering
\caption{Simulation results for the estimated treatment effects for competing risks data using nonparametric estimation} \label{tab:simu1_bias}
\begin{tabular}{lrrrrrr}
  \toprule
 Time & 1 & 2 & 3 & 4 & 5 & 6 \\ 
  \midrule
  \multicolumn{7}{l}{Bias of estimated treatment effects} \\
  Treatment policy & 0.001 & -0.000 & 0.001 & -0.000 & 0.000 & -0.001 \\ 
  Composite variable & 0.000 & -0.000 & 0.001 & 0.000 & 0.001 & -0.001 \\ 
  While on treatment & 0.001 & 0.000 & 0.001 & -0.000 & -0.000 & -0.002 \\ 
  Hypothetical I (natural) & 0.001 & 0.000 & 0.001 & -0.000 & 0.000 & -0.002 \\ 
  Hypothetical II (removed) & 0.001 & 0.000 & 0.001 & -0.000 & 0.001 & -0.001 \\ 
  Principal stratum & 0.001 & 0.001 & 0.001 & -0.002 & -0.002 & -0.003 \\ 
  \midrule
   \multicolumn{7}{l}{Standard deviation of estimated treatment effects} \\
  Treatment policy & 0.013 & 0.024 & 0.032 & 0.040 & 0.045 & 0.050 \\ 
  Composite variable & 0.021 & 0.032 & 0.038 & 0.042 & 0.045 & 0.049 \\ 
  While on treatment & 0.012 & 0.023 & 0.031 & 0.037 & 0.042 & 0.047 \\ 
  Hypothetical I (natural) & 0.012 & 0.023 & 0.030 & 0.036 & 0.040 & 0.045 \\ 
  Hypothetical II (removed) & 0.013 & 0.024 & 0.033 & 0.041 & 0.047 & 0.055 \\ 
  Principal stratum & 0.015 & 0.029 & 0.039 & 0.046 & 0.050 & 0.056 \\ 
   \midrule
   \multicolumn{7}{l}{Mean asymptotic standard error of estimated treatment effects} \\
  Treatment policy & 0.012 & 0.024 & 0.033 & 0.039 & 0.044 & 0.050 \\ 
  Composite variable & 0.022 & 0.032 & 0.040 & 0.043 & 0.046 & 0.049 \\ 
  While on treatment & 0.012 & 0.023 & 0.032 & 0.038 & 0.043 & 0.049 \\ 
  Hypothetical I (natural) & 0.012 & 0.023 & 0.031 & 0.037 & 0.041 & 0.046 \\ 
  Hypothetical II (removed) & 0.012 & 0.024 & 0.034 & 0.042 & 0.048 & 0.055 \\ 
  Principal stratum & 0.015 & 0.029 & 0.039 & 0.046 & 0.052 & 0.057 \\ 
  \midrule
  \multicolumn{7}{l}{Coverage percentage of asymptotic confidence intervals} \\
  Treatment policy & 0.944 & 0.948 & 0.945 & 0.953 & 0.956 & 0.946 \\ 
  Composite variable & 0.949 & 0.949 & 0.958 & 0.950 & 0.961 & 0.951 \\ 
  While on treatment & 0.947 & 0.954 & 0.951 & 0.965 & 0.956 & 0.958 \\ 
  Hypothetical I (natural) & 0.947 & 0.952 & 0.957 & 0.963 & 0.960 & 0.955 \\ 
  Hypothetical II (removed) & 0.944 & 0.951 & 0.952 & 0.963 & 0.959 & 0.954 \\ 
  Principal stratum & 0.945 & 0.950 & 0.952 & 0.957 & 0.954 & 0.959 \\
   \midrule
   \multicolumn{7}{l}{Mean bootstrap standard error of estimated treatment effects} \\
  Treatment policy & 0.012 & 0.024 & 0.033 & 0.039 & 0.044 & 0.049 \\ 
  Composite variable & 0.021 & 0.032 & 0.039 & 0.043 & 0.046 & 0.048 \\ 
  While on treatment & 0.012 & 0.023 & 0.031 & 0.037 & 0.042 & 0.047 \\ 
  Hypothetical I (natural) & 0.012 & 0.023 & 0.031 & 0.036 & 0.040 & 0.044 \\ 
  Hypothetical II (removed) & 0.012 & 0.024 & 0.034 & 0.042 & 0.048 & 0.055 \\ 
  Principal stratum & 0.015 & 0.029 & 0.039 & 0.046 & 0.051 & 0.056 \\ 
  \midrule
  \multicolumn{7}{l}{Coverage percentage of bootstrap confidence intervals} \\
  Treatment policy & 0.942 & 0.947 & 0.949 & 0.951 & 0.954 & 0.949 \\ 
  Composite variable & 0.945 & 0.946 & 0.954 & 0.948 & 0.956 & 0.947 \\ 
  While on treatment & 0.938 & 0.946 & 0.949 & 0.962 & 0.951 & 0.952 \\ 
  Hypothetical I (natural) & 0.938 & 0.947 & 0.952 & 0.961 & 0.950 & 0.940 \\ 
  Hypothetical II (removed) & 0.939 & 0.947 & 0.949 & 0.962 & 0.954 & 0.952 \\ 
  Principal stratum & 0.940 & 0.949 & 0.953 & 0.954 & 0.949 & 0.954 \\ 
   \bottomrule
\end{tabular}
\end{table}

Next, suppose that we collect data in the semi-competing risks data structure. Table \ref{tab:simu2} lists the explicit forms of $\mu_w^k(t)$. Since we have information about the primary outcome event after the intercurrent events, the cumulative incidence function of the primary outcome event under the hypothetical strategy I is larger than that obtained from competing risks data.

\begin{table}
\centering
\caption{The explicit form of the cumulative incidence functions in the simulation with competing risks data} \label{tab:simu2}
\begin{tabular}{ll}
\toprule
Strategy & $\mu_w^k (t)$, $0 \le t \le t^*$ \\ \midrule
tp & $1 - e^{-a_w t^2/2}$ \\
cv & $1 - e^{-a_w t^2/2 - c_w t}$ \\
wo & $1-e^{-a_w t^2/2 - c_w t} - e^{c_w^2/2a_w} (2\pi c_w^2/a_w)^{1/2}\{\Phi(a_w^{1/2}(t+c_w/a_w)) - \Phi(c_w/a_w^{1/2})\}$ \\
hp,I & $1 - e^{-a_w t^2/2}$ \\
hp,II & $1 - e^{-a_w t^2/2}$ \\
ps & $\frac{1-e^{-a_w t^2/2 - c_w t} - e^{c_w^2/2a_w} (2\pi c_w^2/a_w)^{1/2}\{\Phi(a_w^{1/2}(t+c_w/a_w)) - \Phi(c_w/a_w^{1/2})\}}{1 - e^{c_w^2/2a_w} {(2\pi c_w^2/a_w)}^{1/2}\{\Phi(a_w^{1/2}(t^*+c_w/a_w)) - \Phi(c_w/a_w^{1/2})\}}$ \\
\bottomrule
\end{tabular} \\
Notes: $\Phi(\cdot)$ denotes the cumulative distribution function of the standard normal distribution.
\end{table}

Consider a conditional randomized trial (observational study) with two covariates. We simulate data $B=1000$ times and evaluate the empirical performance of semiparametrically efficient estimation. 
In each replicate, we independently generate two covariates from a Bournulli distribution \code{rbinom(n, 1, 0.5)}. The treatment assignment depends on covariates whose probability follows a logistic function: $P(W=1 \mid X) = 1/\{1+\exp(0.5 - 0.3X_1 - 0.6X_2)\}$. The hazards depend on covariates.
We use \code{rweibull(n, 2, sqrt(2 / ax))} to generate the time to the primary outcome event and use \code{rexp(N, cx)} to generate the time to the intercurrent event, where \code{ax} reprresents $\{a_1(x), a_0(x)\}$ and \code{cx} represents $\{c_1(x), c_0(x)\}$. Censoring is generated from the minimum of a uniform distribution \code{runif(n, 4, 8)} and 7. 

\begin{CodeChunk}
\begin{CodeInput}
R> generatedata <- function(n){
R>   X1 = rbinom(n, 1, 0.5) 
R>   X2 = rbinom(n, 1, 0.5) 
R>   X = cbind(X1, X2)
R>   A = rbinom(n, 1, 1 / (1 + exp(0.5 - 0.3 * X1 - 0.6 * X2)))
R>   a1x = a1 + X1 * 0.02 - X2 * 0.03
R>   a0x = a0 + X1 * 0.02 - X2 * 0.01
R>   c1x = c1 + X1 * 0.01 - X2 * 0.02
R>   c0x = c0 - X1 * 0.01 - X2 * 0.01
R>   T1 = rweibull(n, 2, sqrt(2 / a1x))
R>   T0 = rweibull(n, 2, sqrt(2 / a0x))
R>   R1 = rexp(n, c1x)
R>   R0 = rexp(n, c0x)
R>   C = runif(n, 4, 8)
R>   C[C > 7] = 7
R>   T = T1 * A + T0 * (1 - A)
R>   R = R1 * A + R0 * (1 - A)
R>   R[R >= T] = 99
R>   dT = as.numeric(T <= C)
R>   dR = as.numeric(R <= C)
R>   T = T * dT + C * (1 - dT)
R>   R = R * dR + C * (1 - dR)
R>   Time = (T + R - abs(T - R)) / 2
R>   cstatus = dT + 2 * dR
R>   cstatus[cstatus > 2] = 2
R>   return(list(A = A, T = T, R = R, dT = dT, dR = dR, 
R>          Time = Time, cstatus = cstatus, X = X))
R> }
\end{CodeInput}
\end{CodeChunk}

In this simulation, \code{A}, \code{T}, \code{dT}, \code{R}, \code{dR}, and \code{X} from the output of \fct{generatedata} will be used. The true CIF is calculated by averaging the CIFs in \ref{tab:simu2} over the distribution of covariates.
To ease replication of the results, we evaluate the asymptotic standard error (confidence
interval) and bootstrap standard error (confidence interval) from 200 resamples separately. Finishing the simulation for the asymptotic standard error (confidence interval) takes less than an hour, whereas bootstrapping takes several days. Table \ref{tab:simu2_bias} shows the empirical bias of estimated treatment effects, standard deviation of estimated treatment effects among $B=1000$ replicates, mean standard error of estimated treatment effects, and coverage percentage of 95\% confidence intervals at selected time points. The bias is negligible. The standard error is close to the standard deviation. The confidence intervals have coverage rates close to the nominal level. The asymptotic standard errors (confidence intervals) are similar to bootstrap standard errors (confidence intervals), suggesting that using the asymptotic formula is statistically valid and computationally efficient for inference.

\begin{table}
\centering
\caption{Bias of the estimated treatment effects for semi-competing risks data using semiparametrically efficient estimation} \label{tab:simu2_bias}
\begin{tabular}{lrrrrrr}
  \toprule
 Time & 1 & 2 & 3 & 4 & 5 & 6 \\ 
  \midrule
  \multicolumn{7}{l}{Bias of estimated treatment effects} \\
  Treatment policy & 0.000 & 0.001 & -0.001 & -0.001 & -0.002 & -0.003 \\ 
  Composite variable & 0.001 & 0.002 & 0.000 & -0.001 & -0.001 & -0.001 \\ 
  While on treatment & 0.011 & 0.011 & 0.010 & 0.010 & 0.009 & 0.009 \\ 
  Hypothetical I (natural) & 0.000 & 0.001 & -0.001 & -0.002 & -0.002 & -0.004 \\ 
  Hypothetical II (removed) & 0.000 & 0.001 & -0.000 & -0.001 & -0.002 & -0.002 \\ 
  Principal stratum & 0.000 & -0.000 & -0.004 & -0.009 & -0.012 & -0.014 \\ 
  \midrule
   \multicolumn{7}{l}{Standard deviation of estimated treatment effects} \\
  Treatment policy & 0.013 & 0.023 & 0.033 & 0.039 & 0.045 & 0.050 \\ 
  Composite variable & 0.021 & 0.032 & 0.039 & 0.042 & 0.045 & 0.049 \\ 
  While on treatment & 0.013 & 0.023 & 0.032 & 0.038 & 0.043 & 0.050 \\ 
  Hypothetical I (natural) & 0.013 & 0.023 & 0.033 & 0.039 & 0.045 & 0.050 \\ 
  Hypothetical II (removed) & 0.013 & 0.024 & 0.034 & 0.041 & 0.047 & 0.055 \\ 
  Principal stratum & 0.015 & 0.027 & 0.038 & 0.044 & 0.050 & 0.056 \\ 
   \midrule
   \multicolumn{7}{l}{Mean asymptotic standard error of estimated treatment effects} \\
  Treatment policy & 0.013 & 0.024 & 0.033 & 0.040 & 0.045 & 0.050 \\ 
  Composite variable & 0.021 & 0.032 & 0.039 & 0.043 & 0.046 & 0.049 \\ 
  While on treatment & 0.012 & 0.024 & 0.032 & 0.039 & 0.043 & 0.049 \\ 
  Hypothetical I (natural) & 0.013 & 0.024 & 0.033 & 0.040 & 0.045 & 0.050 \\ 
  Hypothetical II (removed) & 0.013 & 0.025 & 0.035 & 0.042 & 0.048 & 0.054 \\ 
  Principal stratum & 0.015 & 0.028 & 0.039 & 0.045 & 0.050 & 0.055 \\ 
  \midrule
  \multicolumn{7}{l}{Coverage percentage of asymptotic confidence intervals} \\
  Treatment policy & 0.958 & 0.958 & 0.963 & 0.958 & 0.955 & 0.945 \\ 
  Composite variable & 0.944 & 0.943 & 0.954 & 0.959 & 0.956 & 0.953 \\ 
  While on treatment & 0.864 & 0.932 & 0.942 & 0.948 & 0.944 & 0.932 \\ 
  Hypothetical I (natural) & 0.957 & 0.952 & 0.961 & 0.957 & 0.957 & 0.943 \\ 
  Hypothetical II (removed) & 0.955 & 0.956 & 0.964 & 0.960 & 0.955 & 0.939 \\ 
  Principal stratum & 0.955 & 0.955 & 0.962 & 0.958 & 0.944 & 0.944 \\
   \midrule
   \multicolumn{7}{l}{Mean bootstrap standard error of estimated treatment effects} \\
  Treatment policy & 0.013 & 0.024 & 0.033 & 0.039 & 0.044 & 0.049 \\ 
  Composite variable & 0.021 & 0.031 & 0.039 & 0.043 & 0.045 & 0.048 \\ 
  While on treatment & 0.012 & 0.024 & 0.032 & 0.039 & 0.044 & 0.049 \\ 
  Hypothetical I (natural) & 0.012 & 0.024 & 0.033 & 0.040 & 0.045 & 0.050 \\ 
  Hypothetical II (removed) & 0.013 & 0.025 & 0.034 & 0.042 & 0.047 & 0.054 \\ 
  Principal stratum & 0.015 & 0.028 & 0.038 & 0.045 & 0.050 & 0.055 \\
  \midrule
  \multicolumn{7}{l}{Coverage percentage of bootstrap confidence intervals} \\
  Treatment policy & 0.955 & 0.950 & 0.958 & 0.957 & 0.952 & 0.937 \\ 
  Composite variable & 0.944 & 0.944 & 0.954 & 0.961 & 0.953 & 0.950 \\ 
  While on treatment & 0.863 & 0.933 & 0.947 & 0.951 & 0.947 & 0.940 \\ 
  Hypothetical I (natural) & 0.955 & 0.952 & 0.961 & 0.958 & 0.950 & 0.937 \\ 
  Hypothetical II (removed) & 0.954 & 0.952 & 0.958 & 0.957 & 0.951 & 0.941 \\ 
  Principal stratum & 0.953 & 0.949 & 0.965 & 0.951 & 0.940 & 0.940 \\ 
   \bottomrule
\end{tabular}
\end{table}

In summary, the simulation studies verify the validity and accuracy of our programs. Considering the extremely high computational burden of bootstrapping, we suggest users use the asymptotic standard error.

\clearpage
\end{appendix}

\end{document}